\RequirePackage{fix-cm}

\documentclass[smallextended]{svjour3}       
\smartqed  

\usepackage{enumitem}
\usepackage{listings}
\usepackage{multirow}
\usepackage[htt]{hyphenat}
\usepackage{makecell}
\usepackage{tcolorbox}

\usepackage{tikz}
\usetikzlibrary{shapes, arrows, positioning}
\tikzstyle{startstop} = [rectangle, rounded corners, minimum height=1cm, text width=4cm, text centered, draw=black, fill=red!30]
\tikzstyle{process} = [rectangle, minimum height=1cm, text width=4cm, text centered, draw=black, fill=orange!30]
\tikzstyle{arrow} = [thick, ->, >=stealth]

\usepackage{pgfplots}
\usepackage{booktabs}
\usepackage{graphicx}
\pgfplotsset{compat=1.17}

\usepackage{threeparttable}

\usepackage[utf8]{inputenc}
\usepackage{tabularx}
\usepackage{siunitx} 

\newcommand{\ie}{\emph{i.e.,}\xspace}

\usepackage{graphicx}
\usepackage{amsmath}
\usepackage{listings}
\usepackage{mdframed}
\usepackage{multirow}
\usepackage{tikz}
\usetikzlibrary{arrows.meta, calc, shapes}
\usepackage{makecell}
\usepackage{subcaption}
\usepackage{rotating}
\usepackage{tcolorbox}
\usepackage{enumitem}
\usepackage{adjustbox}
\usepackage{xspace}
\usepackage{xcolor}
\usepackage{pifont}
\usepackage{booktabs}
\usepackage{url}
\usepackage[colorlinks=true,linkcolor=blue,citecolor=blue,urlcolor=blue]{hyperref}
\usepackage{natbib}

\definecolor{codegreen}{rgb}{0,0.6,0}
\definecolor{codegray}{rgb}{0.5,0.5,0.5}
\definecolor{codepurple}{rgb}{0.58,0,0.82}
\definecolor{codeblue}{rgb}{0.0, 0.0, 0.55}

\journalname{Empirical Software Engineering}

\begin{document}

\title{Beyond Synthetic Benchmarks: Evaluating LLM Performance on Real-World Class-Level Python Code Generation}

\author{Musfiqur Rahman \and 
        SayedHassan Khatoonabadi \and 
        Emad Shihab}

\institute{Musfiqur Rahman \at
           Concordia University, Montr\'{e}al, QC, Canada \\
           \email{musfiqur.rahman@mail.concordia.ca}
           \and
           SayedHassan Khatoonabadi \at
           Concordia University, Montr\'{e}al, QC, Canada \\
           \email{sayedhassan.khatoonabadi@concordia.ca}
           \and
           Emad Shihab \at
           Concordia University, Montr\'{e}al, QC, Canada \\
           \email{emad.shihab@concordia.ca}}

\date{Received: date / Accepted: date}

\maketitle

\begin{abstract}
Large language models (LLMs) have advanced rapidly from well-studied function-level code generation to emerging repository-level approaches, yet this progression reveals a critical gap: class-level code generation, which is the fundamental building block connecting individual functions to complete repositories, remains systematically underexplored. Understanding class-level capabilities is a prerequisite to repository-level success, as current approaches cannot distinguish whether failures stem from basic implementation deficiencies or sophisticated context management issues.

We introduce \textit{RealClassEval}, a benchmark derived from real-world open-source repositories, comprising 400 classes divided into seen and unseen partitions based on model training cutoffs. Our systematic evaluation across seven state-of-the-art LLMs reveals fundamental limitations masked by synthetic benchmarks: while LLMs achieve 84–89\% correctness on synthetic tasks, they attain only 25–34\% on real-world classes, with minimal distinction between seen and unseen codebases.

Notably, this performance range converges with concurrent work using rich natural language descriptions and repository context, suggesting that specification complexity cannot overcome core implementation bottlenecks. Documentation provides marginal improvements (1–3\%), while retrieval augmentation yields greater gains (4–7\%,) specifically with incomplete specifications. Error analysis reveals AttributeError, TypeError, and AssertionError dominate 84\% of failures, with distinct patterns between synthetic (logic-focused) and real-world (object-oriented) scenarios.

These findings establish the diagnostic foundation necessary for advancing both fundamental LLM capabilities and sophisticated repository-level approaches, while providing actionable guidance for production deployment strategies.

\end{abstract}

\section{Introduction}
\label{sec:introduction}

Large language models (LLMs) have demonstrated remarkable progress in code generation, enabling tools like GitHub Copilot~\citep{githubGitHubCopilot} and Amazon CodeWhisperer~\citep{amazonCodeWhispererBecoming} to assist developers across multiple granularities~\citep{chen2021evaluating,tao2025retrieval}. The field has evolved rapidly from well-studied function-level generation~\citep{chen2021evaluating,austin2021program} to emerging repository-level approaches~\citep{jimenez2023swe,zheng2024humanevo,zhang2024codeagent} that tackle complex multi-file scenarios. However, this progression reveals a critical gap: \textbf{class-level code generation}, which is the fundamental building block connecting individual functions to complete repositories, remains systematically underexplored.

This gap has profound implications for both understanding LLM capabilities and advancing repository-level approaches. Current repository-level benchmarks like SWE-bench~\citep{jimenez2023swe} and recent advances in environment-aware generation~\citep{wu2026environment} implicitly assume successful class-level implementation capabilities. Yet when repository-level approaches fail, we cannot determine whether failures stem from class-level implementation deficiencies, context management issues, or dependency resolution challenges. Without understanding baseline class-level capabilities, researchers cannot effectively focus their architectural innovations or set realistic expectations for repository-level performance.

Existing class-level evaluation relies primarily on \textbf{ClassEval}~\citep{du2023classeval}, which uses manually crafted synthetic tasks that do not reflect real-world complexity. Synthetic benchmarks present three fundamental limitations relative to real-world code. First, synthetic classes are self-contained with minimal dependencies, whereas real-world code requires framework integrations and project-specific patterns. Second, synthetic test suites rely on simplified assertions, whereas production code demands complex type systems and inheritance hierarchies. Third, synthetic benchmarks lack evolutionary complexity, since real-world classes must integrate with existing architectural patterns. These differences create a false sense of LLM capabilities because high performance on synthetic benchmarks provides no guarantee of success on realistic tasks, potentially misleading both researchers developing repository-level approaches and practitioners deploying LLM-based tools.

We address the issues above by curating a \textbf{new benchmark of real-world class-level code} from open-source projects, categorized as \textbf{`seen'} (classes from repositories created before model training cutoffs, likely encountered during training) and \textbf{`unseen'} (classes from repositories created after training cutoffs, guaranteeing they were not in training data). Our evaluation framework assesses LLM performance across different docstring variants, generation strategies, and RAG-augmented setups. We systematically analyze the impact of docstrings, the effectiveness of RAG, and the types of errors that arise in generated code, providing insights into the strengths and limitations of current LLMs. More specifically, we investigate the following research questions:
\begin{itemize}[leftmargin=*]
    \item \textbf{RQ1 (Real-World Performance): How do LLMs perform on class-level code generation in real-world projects compared to synthetic benchmarks?}
    We find that LLMs achieve 84--89\% correctness on synthetic benchmarks but only 24--35\% on real-world code. Models show nearly identical performance on seen and unseen real-world data.
    \item \textbf{RQ2 (Docstring Impact): Does the presence of docstrings improve the functional correctness of LLM-generated classes?}
    Our finding shows that complete docstrings provide improvements of 1--3\% in correctness, in some cases, although this improvement is not frequent enough to gain statistical significance. 
    \item \textbf{RQ3 (RAG for Unseen Code): Can retrieval-augmented generation (RAG) using seen data improve LLM performance on unseen class-level tasks?}
    We find that RAG provides substantial benefits of 4--7\% specifically when documentation is incomplete (partial docstrings), but minimal value with complete documentation. This shows that RAG helps by providing concrete implementation examples when specifications lack implementation details.
    \item \textbf{RQ4 (Error Analysis): What are the most common errors in LLM-generated class-level code?}
    We find that \texttt{AttributeError}, \texttt{TypeError}, and \texttt{AssertionError} account for 84\% of all failures. Synthetic benchmarks show assertion-dominated failures while real-world code shows attribute-and-type-dominated failures, and RAG operates through error substitution, reducing logic failures while occasionally introducing dependency errors.
\end{itemize}

By answering the above RQs, this paper makes the following contributions:
\begin{enumerate}[leftmargin=*]
    \item A curated dataset of \textbf{real-world class-level code} from open-source projects, categorized as seen and unseen.
    \item A systematic evaluation of LLM performance on class-level generation, comparing \textbf{real-world projects} to synthetic benchmarks.
    \item An analysis of the impact of \textbf{docstrings} and \textbf{RAG} on functional correctness and error types.
    \item Insights into \textbf{common errors} and practical implications for improving LLM-assisted programming tools.
\end{enumerate}

The remainder of this paper is organized as follows: Section~\ref{sec:background} discusses related work and background. Section~\ref{sec:methodology} describes the design of our benchmark and the overall view of the research methodology. Sections~\ref{sec:rq1}, \ref{sec:rq2}, \ref{sec:rq3}, and \ref{sec:rq4} address RQs 1, 2, 3, and 4, respectively. Section~\ref{sec:discussion} discusses the implications of our findings. Section~\ref{sec:limitations} describes some of the limitations of our study and outlines future directions. Section~\ref{sec:threats} discusses the threats to validity. Finally, Section~\ref{sec:conclusion} concludes the paper.

\section{Background and Related Work}
\label{sec:background}

The evolution of code generation benchmarks follows a clear progression from function-level to repository-level tasks, yet a critical gap exists at the class-level. This section positions our contribution within this progression and establishes why class-level evaluation represents a necessary prerequisite for understanding repository-level capabilities.

\subsection{Function-Level Code Generation}
Large Language Models (LLMs) have achieved remarkable success in function-level code generation, enabling tools like GitHub Copilot and Amazon CodeWhisperer to assist developers~\citep{chen2021evaluating,austin2021program}. Established benchmarks like \textbf{HumanEval}~\citep{chen2021evaluating} and \textbf{MBPP}~\citep{austin2021program} evaluate standalone function generation, while \textbf{CoderEval}~\citep{yu2024codereval} introduces non-standalone functions but maintains function-level granularity.

To understand current limitations, \textbf{Wang et al.}~\citep{wang2025towards} analyzed 557 incorrect code snippets from six prominent LLMs on HumanEval, finding that syntactic errors account for 40\% of failures, with a clear correlation between model size and error frequency. \textbf{Liu et al.}~\citep{liu2023your} introduced \textbf{EvalPlus}, which revealed that HumanEval's test insufficiency allows models to pass with fragile code, reducing pass@k scores by 19.3--28.9\% when comprehensive test suites are applied. These findings demonstrate that while LLMs excel at isolated function generation, they struggle with structural correctness and comprehensive test coverage.

\subsection{Repository-Level Code Generation: The Emerging Frontier}
Repository-level code generation has attracted growing attention as the ultimate goal for practical software engineering automation. Recent work addresses the complexities of multi-file, cross-dependency code generation through various approaches.

\textbf{SWE-bench}~\citep{jimenez2023swe} introduced repository-level issue resolution, requiring models to understand and modify existing codebases to fix real GitHub issues. \textbf{HumanEvo}~\citep{zheng2024humanevo} extends this with evolution-aware evaluation, tracking how repository structure changes over time. \textbf{CodeAgent}~\citep{zhang2024codeagent} proposes tool-integrated agent systems for repository-level coding challenges, demonstrating the need for sophisticated architectural scaffolding. For example, repository-level benchmarks like \textbf{RepoExec}~\citep{le2024repoexec} evaluate executability and cross-file context utilization, while surveys by \textbf{Tao et al.}~\citep{tao2025retrieval} systematically categorize repository-level approaches and their limitations.

Recent advances continue pushing repository-level boundaries. \textbf{Wu et al.}~\citep{wu2026environment} introduced \textbf{Environment-Aware Code Generation}, where generated code must be functionally correct and executable under arbitrary software configurations, revealing that current LLMs struggle with environment-specific requirements. \textbf{InlineCoder}~\citep{hu2026line} enhances repository understanding by inlining unfinished functions into call graphs, reframing repository challenges as function-level tasks. \textbf{CodeMEM}~\citep{wang2026codemem} addresses memory management in interactive repository-level generation through AST-guided dynamic context preservation.

\subsection{Class-Level Code Generation: The Critical Gap}
Between function-level and repository-level generation lies class-level code generation, which is a critical granularity that has received limited systematic evaluation. \textbf{ClassEval}~\citep{du2023classeval} represents the primary evaluation in this space, providing 100 manually crafted class-level tasks. However, its synthetic nature and small scale limit understanding of real-world class generation capabilities.

\textbf{JavaBench}~\citep{cao2024javabench} evaluates object-oriented code generation but focuses on project-level tasks requiring multiple interdependent classes rather than isolated class implementation, using data derived from university course projects rather than real-world software development. While valuable, this conflates class-level implementation capabilities with repository-level dependency resolution and does not reflect the complexity and patterns found in production codebases.

Recent concurrent work introduced \textbf{RepoClassBench}~\citep{deshpande2024class}, which evaluates class-level generation from natural language descriptions within repository contexts. RepoClassBench emphasizes cross-file dependencies and repository-level reasoning, proposing the Retrieve-Repotools-Reflect (RRR) framework that equips LLMs with static analysis tools for iterative repository navigation. While RepoClassBench addresses repository-level complexity, it focuses on natural language-to-code generation with extensive tool augmentation rather than assessing fundamental class implementation capabilities from structured specifications.

\subsection{Retrieval-Augmented Generation for Code}
Retrieval-Augmented Generation (RAG) enhances LLM performance by incorporating relevant context during generation~\citep{yang2025empirical,tao2025retrieval}. \textbf{CodeRAG-Bench}~\citep{wang2024coderag} systematically evaluated retrieval effectiveness across basic programming, open-domain, and repository-level tasks, revealing that while high-quality contexts improve generation, current retrieval models struggle with useful context fetching, particularly when queries lack lexical overlap with relevant documents.

\textbf{CodeRAG}~\citep{li2025coderag} addresses repository-level challenges by constructing dual graphs modelling functional dependencies and semantic relationships, achieving substantial improvements on repository-level benchmarks. However, these approaches assume successful implementation of individual components--an assumption our work tests by evaluating whether RAG helps at the fundamental class-level granularity.

\subsection{Docstrings and Prompt Engineering}
Documentation quality significantly impacts code generation performance~\citep{yang2024less,ronanki2025prompt}. \textbf{Yang et al.}~\citep{yang2024less} demonstrated that strategic docstring compression (25--40\% reduction) preserves generation quality while improving efficiency, revealing that not all documentation contributes equally. \textbf{Taherkhani et al.}~\citep{taherkhani2024epic} showed that evolutionary prompt engineering achieves 6\% improvement in pass@k while being 2--10 times more cost-effective than baselines.

However, existing work focuses on function-level tasks where docstrings describe single-purpose functionality. Class-level docstrings must describe complex object behaviours, method interactions, and state management--fundamentally different documentation challenges that require targeted evaluation.

\subsection{Research Gap and Contribution}
Current evaluation practices exhibit a critical discontinuity: function-level generation is well-studied, repository-level generation is rapidly advancing, but class-level generation, which connects these granularities, remains underexplored. This creates a fundamental knowledge gap about LLM capabilities.

\textbf{The Class-Level Prerequisite Argument:} Repository-level generation inherently requires successful class-level implementation. SWE-bench tasks often involve modifying or creating individual classes within the repository context. Coding agents' tasks are not limited to the generation of functions only; they ultimately generate class implementations. Environment-aware generation must ensure classes execute correctly across configurations. Yet existing evaluations cannot isolate whether failures stem from class-level implementation deficiencies or repository-level context management issues.

This gap has practical implications: if LLMs cannot reliably implement individual classes given clear specifications, repository-level approaches may be addressing the wrong problem. Our work fills this gap by systematically evaluating class-level capabilities independent of repository context, providing the foundation for understanding where repository-level approaches should focus their efforts.

Rather than competing with repository-level work, our contribution (detailed in Section~\ref{sec:introduction}) is to supply the prerequisite understanding it depends on: by isolating class-level capabilities from repository context, we enable precise diagnosis of where current LLMs struggle and where future repository-level approaches should focus their architectural innovations.

\section{Methodology}
\label{sec:methodology}

\subsection{Dataset Curation}
\label{subsec:dataset_curation}

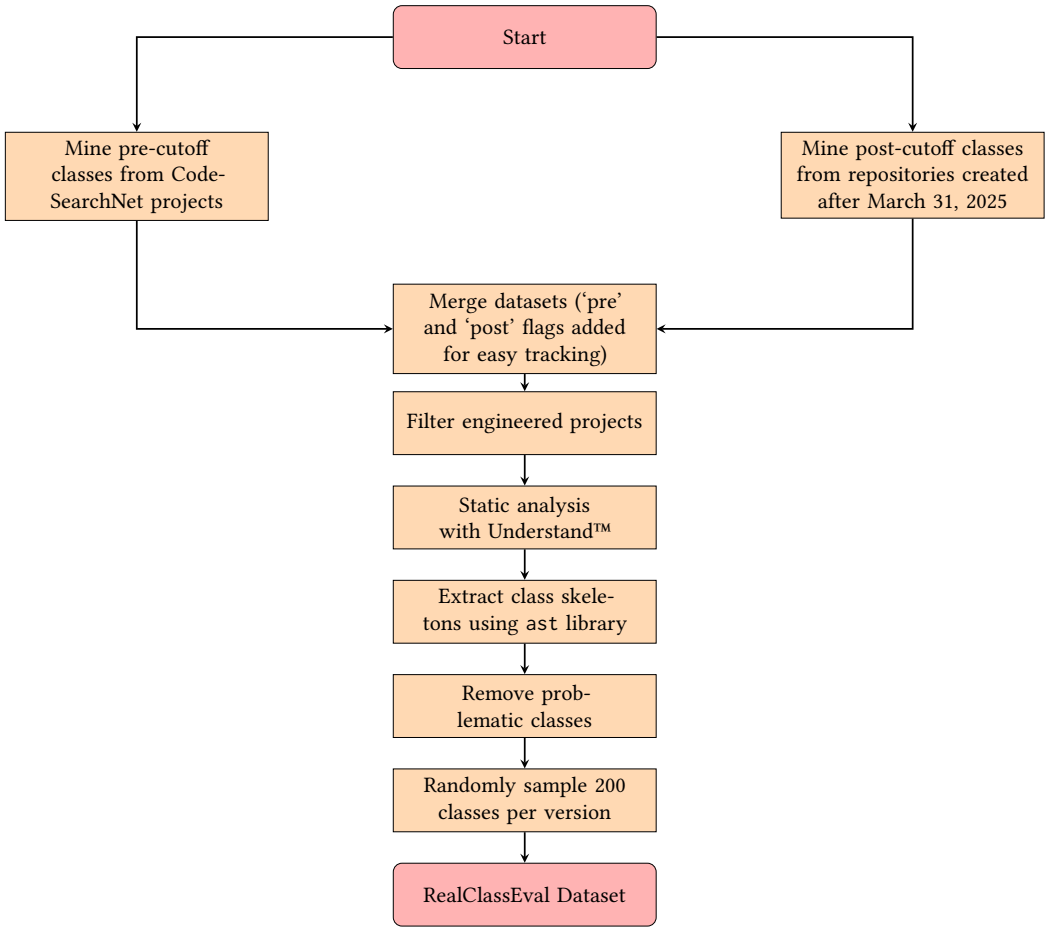
\begin{figure}[h]
    \centering
    \resizebox{\columnwidth}{!}{
    \begin{tikzpicture}[node distance=1.5cm]
        \node (start) [startstop] {Start};
        \node (codesearchnet) [process, below left=1cm and 2cm of start] {Mine pre-cutoff classes from CodeSearchNet projects};
        \node (github) [process, below right=1cm and 2cm of start] {Mine post-cutoff classes from repositories created after March 31, 2025};
        \node (merge) [process, below right=1cm and 2cm of codesearchnet] {Merge datasets (`pre' and `post' flags added for easy tracking)};
        \node (filter) [process, below of=merge] {Filter engineered projects};
        \node (static) [process, below of=filter] {Static analysis with Understand\texttrademark{}};
        \node (ast) [process, below of=static] {Extract class skeletons using \texttt{ast} library};
        \node (postprocess) [process, below of=ast] {Remove problematic classes};
        \node (sample) [process, below of=postprocess] {Randomly sample 200 classes per version};
        \node (end) [startstop, below of=sample] {\textit{RealClassEval} Dataset};

        \draw [arrow] (start) -| (codesearchnet);
        \draw [arrow] (start) -| (github);
        \draw [arrow] (codesearchnet) |- (merge);
        \draw [arrow] (github) |- (merge);
        \draw [arrow] (merge) -- (filter);
        \draw [arrow] (filter) -- (static);
        \draw [arrow] (static) -- (ast);
        \draw [arrow] (ast) -- (postprocess);
        \draw [arrow] (postprocess) -- (sample);
        \draw [arrow] (sample) -- (end);
    \end{tikzpicture}}
    \caption{Workflow for \textit{RealClassEval} dataset curation.}
    \label{fig:dataset_workflow}
\end{figure}

To evaluate the performance of LLMs on real-world class-level code generation, we introduce \textit{RealClassEval}, a dataset comprising two versions of Python classes: \textit{pre-cutoff} and \textit{post-cutoff}. Each version contains 200 classes mined from real-world GitHub open-source projects. The complete data curation process is illustrated in Figure~\ref{fig:dataset_workflow}, which shows our systematic approach from initial data collection through final dataset construction.

\subsubsection{Data Collection:}
The \textit{pre-cutoff} data is sourced from the \textbf{CodeSearchNet} dataset~\citep{husain2019codesearchnet}, which is one of the most widely used datasets for training code-related LLMs. \textbf{CodeSearchNet} has been used as a pre-training corpus for numerous prominent models, including StarCoder~\citep{li2023starcoder}, CodeT5+~\citep{wang2023codet5plus}, and UniXcoder~\citep{guo2022unixcoder}, and has been used in numerous existing research papers~\citep{jiang2024survey,shi2023cocosoda,rahman2024automatic,orel2025codet}. The dataset comprises 2 million (comment, code) pairs from open-source GitHub repositories across six programming languages. Even though we cannot guarantee that this dataset was indeed part of the training data of the LLMs we investigated in this research, we believe that it is a safe assumption to make, given the widespread adoption and popularity of this dataset in the LLM training community. The \textit{post-cutoff} data consists of repositories created after March 31, 2025~\footnote{All LLMs under investigation were either released before this date or their knowledge cut-off dates are before this date.}, guaranteeing these classes are unseen by the evaluated LLMs. 

For the \textit{post-cutoff} data collection, we used the GitHub search API to identify Python repositories created after March 31, 2025 systematically. We selected repositories that: (1) were created after our cutoff date to ensure temporal separation from LLM training data, (2) had active development with at least 10 commits to ensure maturity, (3) contained Python as the primary programming language, and (4) were not forks to eliminate duplicates. This systematic selection process yielded a pool of candidate repositories from which we subsequently applied filtering criteria to identify engineered projects.

\subsubsection{Data Filtering}
To ensure our dataset comprises high-quality, production-level software, we retain only \textit{engineered projects}~\citep{munaiah2017curating} from both the CodeSearchNet dataset and the post-cutoff repositories, adhering to the filtering criteria outlined in~\citep{xiao2025self}. Importantly, we did not include all projects from CodeSearchNet; rather, we applied the same rigorous filtering criteria to both data sources to ensure consistency and quality. Specifically, we applied three filtering criteria to both \textit{pre-cutoff} (CodeSearchNet) and \textit{post-cutoff} (newly collected) repositories: 
\begin{enumerate}
    \item \textbf{License Declaration}: We excluded repositories not declaring a license or using non-standard licenses (marked as ``Other'' in the GitHub search tool). For the remaining repositories, we removed projects declaring licenses not commonly used for software projects, such as Creative Commons Attribution licenses and font licenses, as these typically indicate documentation or non-code projects.
    \item \textbf{Project Size and Activity}: We filtered out repositories with fewer than a minimum threshold of stars (10+) and commits (50+) to ensure the projects represent mature, community-recognized software rather than personal experiments or abandoned prototypes.
    \item \textbf{Code Quality Indicators}: We excluded projects with incomplete documentation, lack of structured code organization (e.g., no clear module structure), or repositories that primarily contain tutorials, educational materials, or code snippets rather than production-quality software.
\end{enumerate}
These filtering criteria, applied uniformly to both CodeSearchNet and post-cutoff repositories, ensure that our dataset comprises classes from well-maintained, professionally developed open-source projects that reflect real-world software engineering practices.

We report the concrete filtering funnel for both datasets in Appendix~\ref{app:filtering_funnel} (Table~\ref{tab:filtering_funnel}).

\subsubsection{Static Analysis and Class Extraction:}
From the filtered engineered projects in both data sources, we use \textbf{Understand\texttrademark{}}~\citep{scitoolsUnderstandSoftware} to perform static analysis and identify Python classes. Class extraction is carried out using Python's \texttt{ast} library, parsing all \texttt{.py} files to extract class skeletons while preserving class and method signatures and docstrings. An example of the class skeleton extraction process is shown in Listings \ref{lst:example_class} and \ref{lst:example_skeleton}. They illustrate how we transform a human-written class into a skeleton that serves as input for LLM code generation.

\begin{lstlisting}[
    language=Python,
    basicstyle=\ttfamily\small,
    breaklines=true,
    caption={An example Python class from \texttt{codenerix/django-codenerix}~\citep{githubGitHubTkaemmingdjangosubdomains}.},
    label=lst:example_class,
    captionpos=t,
    commentstyle=\color{codegray},
    keywordstyle=\color{codeblue}\bfseries,
    stringstyle=\color{codegreen},
    showstringspaces=false,
    showspaces=false,
    numbers=none
]
class SubdomainMiddleware(object):
    """
    A middleware class that adds a ``subdomain'' attribute to the current request.
    """
    def get_domain_for_request(self, request):
        """
        Returns the domain that will be used to identify the subdomain part
        for this request.
        """
        return get_domain()

    def process_request(self, request):
        """
        Adds a ``subdomain'' attribute to the ``request'' parameter.
        """
        domain, host = map(lower,
            (self.get_domain_for_request(request), request.get_host()))

        pattern = r'^(?:(?P<subdomain>.*?)\.)?%s(?::.*)?$' % re.escape(domain)
        matches = re.match(pattern, host)

        if matches:
            request.subdomain = matches.group('subdomain')
        else:
            request.subdomain = None
            logger.warning('The host %s does not belong to the domain %s, '
                'unable to identify the subdomain for this request',
                request.get_host(), domain)
\end{lstlisting}

\begin{lstlisting}[
    language=Python,
    basicstyle=\ttfamily\small,
    breaklines=true,
    caption={\textcolor{blue}{Docstring} extracted from the class shown in Listing~\ref{lst:example_class}.},
    label=lst:example_skeleton,
    captionpos=t,
    commentstyle=\color{codegray},
    keywordstyle=\color{codeblue}\bfseries,
    stringstyle=\color{codegreen},
    showstringspaces=false,
    showspaces=false,
    numbers=none
]

class SubdomainMiddleware(object):
    """
    A middleware class that adds a ``subdomain'' attribute to the current request.
    """

    def get_domain_for_request(self, request):
        """
        Returns the domain that will be used to identify the subdomain part
        for this request.
        """
        pass

    def process_request(self, request):
        """
        Adds a ``subdomain'' attribute to the ``request`` parameter.
        """
        pass

\end{lstlisting}

\subsubsection{Post-processing and Sampling}

Post-processing involves removing problematic classes (e.g., those with Python 2 syntax), ensuring only parsable and valid skeletons remain. Following \textit{ClassEval}'s exclusion criteria~\citep{du2023classeval}, we omit tasks with complex execution dependencies, identified through static metrics. However, unlike \textit{ClassEval}, we do not filter by domain, retaining a diverse range of topics such as Game Development, File Handling, and Management Systems. This approach ensures our dataset reflects the heterogeneity of real-world projects, providing a robust foundation for evaluating LLM performance on class-level code generation. From the remaining classes, we randomly sample 200 classes per version (\textit{pre-cutoff} and \textit{post-cutoff}) to create a balanced dataset for evaluation.

\subsubsection{Dataset Comparability Validation}

To ensure valid temporal comparisons and confirm that observed performance differences stem from model capabilities rather than dataset characteristics, we conducted a comprehensive statistical analysis comparing pre-cutoff and post-cutoff distributions across key dimensions.

Our analysis reveals no statistically significant differences between the datasets across all measured characteristics (all p-values $> 0.05$, Mann-Whitney U tests):

\begin{itemize}[leftmargin=*]
\item \textbf{Code Complexity}: Both datasets show comparable structural complexity (lines of code: 74.3±95.8 vs 187.8±287.0, p=0.259; methods per class: 2.5±1.4 vs 3.7±2.2, p=0.402; cyclomatic complexity: 3.7±5.0 vs 4.0±4.5, p=0.699).
\item \textbf{Documentation Characteristics}: Documentation patterns are statistically indistinguishable (total docstrings: 3.4±1.3 vs 4.5±2.4, p=0.516; comment-to-code ratio: 1.7±2.6 vs 1.2±1.8, p=0.383).
\item \textbf{Object-Oriented Features}: Class design characteristics show no significant differences (instance methods: 2.2±1.6 vs 3.2±2.5, p=0.368; class coupling: 3.8±5.9 vs 5.5±6.2, p=0.224).
\end{itemize}

While post-cutoff classes show numerically higher complexity values, these differences fall within expected random sampling variation and do not reach statistical significance under the tests applied. This comparability across structural, documentation, and design characteristics is consistent with our temporal split methodology and supports treating the pre-cutoff and post-cutoff datasets as comparable populations for experimental purposes. However, we acknowledge that the absence of a detected difference does not by itself establish equivalence. We treat this as a methodological assumption, not a proven guarantee.

These comparable distributions ensure that our seen versus unseen comparisons (RQ1), docstring effects experiments (RQ2), retrieval-augmented generation experiments (RQ3), and error pattern analyses (RQ4) reflect genuine model capabilities rather than confounding dataset characteristics. The statistical equivalence of our temporal splits provides a strong methodological foundation for attributing observed performance differences to the experimental factors under investigation.

\subsection{Code Generation}
\label{sec:code_generation}

\subsubsection{Selected LLMs}
\label{subsec:selected_llms}

To evaluate the performance of LLMs on class-level code generation, we selected seven state-of-the-art models: \textbf{Qwen 2.5 Coder}~\citep{hui2024qwen2}, \textbf{GPT-4.1}~\citep{openaiIntroducingGPT41}, \textbf{GPT-5}~\citep{openaiIntroducingGPT5}, \textbf{GPT-OSS}~\citep{agarwal2025gpt}, \textbf{Codestral}~\citep{mistralCodestralMistral}, \textbf{DeepSeek-V3}~\citep{liu2024deepseek}, and \textbf{Llama-4 Maverick}~\citep{metaLlamaHerd}. These models were chosen to represent a diverse range of architectures, training paradigms, and capabilities, ensuring a comprehensive evaluation of both general-purpose and code-specific LLMs, as well as models with varying reasoning abilities. They also vary in their release dates, knowledge cutoffs, parameter sizes, context window lengths, and specialization. Table~\ref{tab:llm_characteristics} summarizes these characteristics and highlights the diversity of our evaluation set.

\begin{table*}[t]
    \centering
    \caption{Characteristics of the evaluated LLMs.}
    \label{tab:llm_characteristics}
    \resizebox{\columnwidth}{!}{
    \begin{tabular}{|l|c|c|c|p{4cm}|}
        \hline
        \textbf{Model} & \textbf{Release Date} & \textbf{Knowledge Cutoff} & \textbf{Parameters} & \textbf{Key Characteristics} \\ \hline
        Qwen 2.5 Coder & September 2024 & June 2024 & 32B & Code-specific, optimized for code generation and reasoning \\ \hline
        GPT-4.1 & April 2025 & June 2024 & 1.8T (estimated) & General-purpose, advanced reasoning and instruction-following \\ \hline
        GPT-5 & August 2025 & October 2024 & 2.5T (estimated) & General-purpose, state-of-the-art reasoning and context handling \\ \hline
        GPT-OSS & August 2025 & June 2024 & 20B & General-purpose (open), strong code capabilities \\ \hline
        Codestral & January 2025 (Latest) & Unknown & 100B & Code-specific, specialized for code, lightweight and efficient \\ \hline
        DeepSeek-V3 & August 2025 & July 2024 & 126B & General-purpose, focused on long-context code understanding \\ \hline
        Llama-4 Maverick & April 2025 & August 2024 & 400B & General-purpose, balanced reasoning and code generation \\ \hline
    \end{tabular}}
\end{table*}

\subsubsection{Prompt Design}
\label{subsec:prompt_design}

We use a standardized prompt to ensure consistency across all models. The prompt is designed to instruct the LLM to implement a complete Python class based on the provided class skeleton, without additional explanations. The prompt template is as follows:

\noindent
\fbox{
    \parbox{\columnwidth}{
        \small You are an expert Python programmer who can correctly implement complete Python classes based on the provided class skeletons. Implement the following class. Do not explain the code. The given class skeleton is as follows: \newline
        \small[CLASS SKELETON]
    }
}

Unlike \textit{ClassEval}~\citep{du2023classeval}, which explores three generation strategies (holistic, incremental, and interactive), we focus exclusively on \textbf{holistic generation}. This decision is motivated by the findings of ClassEval, which indicate that models with longer context windows (such as GPT-3.5 and GPT-4 in their study) perform well in holistic generation. Given that all our selected models have substantial context windows (ranging from 128K to 1M tokens), we opt for holistic generation to leverage their ability to process and generate entire class implementations in a single pass.

\subsection{Test Suite Generation}
\label{subsec:test_suite_generation}

To assess functional correctness, we use \textbf{PYNGUIN}~\citep{lukasczyk2022pynguin} to automatically generate test suites for the classes in our dataset. \textbf{PYNGUIN} (the PYthoN General UnIt test geNerator) is an automated unit test generation framework specifically designed for Python programs. Unlike test generation tools for statically typed languages like Java, \textbf{PYNGUIN} addresses the unique challenges posed by Python's dynamic typing and lack of explicit type information. The tool takes a Python module as input along with its dependencies and automatically generates unit tests that maximize code coverage using various test generation algorithms.

We specifically employ \textbf{PYNGUIN}'s \textbf{DynaMOSA} (Dynamic Many-Objective Sorting Algorithm)~\citep{panichella2017automated} algorithm for test generation. \textbf{DynaMOSA} is a many-objective evolutionary algorithm specifically designed for automated test case generation with dynamic selection of coverage targets. The algorithm treats test case generation as a many-objective optimization problem where each coverage target (e.g., branches, statements) is treated as a separate objective. \textbf{DynaMOSA} dynamically adjusts its focus based on the control dependency hierarchy, prioritizing uncovered targets and progressively exploring the search space. In the original paper that introduced PYNGUIN, DynaMOSA achieved 68.0\% mean branch coverage across their benchmark, outperforming other search-based algorithms (MIO: 67.0\%, MOSA: 67.8\%) and significantly exceeding random generation (63.6\%), establishing it as the most effective approach for Python test generation~\citep{lukasczyk2022pynguin}.

\textbf{PYNGUIN invocation parameters.} We used PYNGUIN version 0.40.0 with the following configuration, applied identically to every class: maximum iterations 100; maximum search time 180 seconds per class (whichever limit is reached first terminates the search); assertion generation mode \texttt{SIMPLE}; and a single fixed random seed, $175$, applied uniformly across all classes. Branch coverage was extracted directly from PYNGUIN's own per-run coverage report.

\subsubsection{Coverage Analysis and Methodological Validation}
\label{subsec:coverage_analysis}

Our coverage analysis reveals that PYNGUIN achieves markedly different coverage patterns across synthetic versus real-world code: 93.2\% average branch coverage on ClassEval's manually crafted classes, versus 58.7\% on \textit{RealClassEval} classes. Both figures are consistent with DynaMOSA's reported 68.0\% mean coverage in its original evaluation~\citep{lukasczyk2022pynguin}, and the differential tracks the added structural complexity of real-world code, which includes deeper conditional logic, broader error handling, and external integration points, relative to synthetic benchmarks' simplified control flow. This pattern indicates PYNGUIN is sensitive to the same implementation complexity this study investigates. We discuss the measurement limitations this coverage gap introduces in Section~\ref{sec:threats}. To assess whether generated test suites include explicit correctness checks rather than relying solely on implicit exception-based checks, we sampled $n=78$ \textit{RealClassEval} classes -- the size required for a 10\% margin of error at 95\% confidence given our population of 400 real-world classes~\citep{cochran1977sampling}. These classes were allocated proportionally across a 2$\times$2 stratification by pre-cutoff/post-cutoff origin and pass/fail outcome (25 pre-cutoff-fail, 12 pre-cutoff-pass, 27 post-cutoff-fail, 14 post-cutoff-pass). We coded each class's generated test suite as \textit{assertion-based} (containing at least one explicit \texttt{assert} statement, whether checking type, state, or return values) or \textit{assertion-free} (relying solely on \texttt{pytest.mark.xfail}-decorated calls that implicitly expect an exception without any explicit check). 69 of 78 (88.5\%) test suites were assertion-based; 9 (11.5\%) were assertion-free. This assessment establishes that the large majority of generated test suites contain explicit assertions, addressing whether tests check anything beyond exception-free execution. It does not, however, establish that those assertions correctly capture the intended semantics of the original class, or how effectively the suites would detect functional faults (e.g., via mutation testing). We treat semantic and fault-detection adequacy of the automatically generated oracle as an open limitation, discussed further in Section~\ref{sec:limitations}.

We adopt PYNGUIN over alternative test-construction approaches for scalability and consistency. Building executable environments for hundreds of unmaintained real-world repositories to run their native test suites is intractable at our scale. Manual test authoring for 400 diverse classes would similarly require months of expert effort while introducing subjective bias. PYNGUIN instead applies identical algorithmic parameters across all classes and models, enabling directly comparable evaluation across the full dataset without manual intervention.

\subsection{Result Analysis}
\label{subsec:result_analysis}

\subsubsection{Performance Analysis}
\label{subsubsec:performance_analysis}
To evaluate the performance of LLMs on class-level code generation, we employ \textbf{pass rate} as our primary metric, defined as the ratio of passed tests to the total number of tests for each generated class. This metric is directly related to the widely used \textbf{Pass@1}~\citep{chen2021evaluating}, which measures the probability that a generated solution passes all tests in a single attempt. While \textbf{Pass@1} is typically used for function-level evaluation, our pass rate extends this concept to class-level correctness, providing a robust measure of functional accuracy across the entire class implementation. For a given class \(i\), the pass rate metric is defined as:
\begin{align}
\text{Pass Rate}_i = \frac{\text{Number of Passed Tests}_i}{\text{Total Number of Tests}_i}
\label{eq:pass_rate}
\end{align}

We compare one pass rate distribution against another to see:
\begin{itemize}
    \item if a statistically significant difference exists between distributions.
    \item when a statistically significant difference exists, how big the difference is in terms of effect size.
\end{itemize}
We detail the statistical approaches taken to answer the RQs in their respective sections.

\subsubsection{Error Analysis:}
\label{subsubsec:error_analysis}
We analyze the distribution of error types (e.g., \texttt{AttributeError}, \texttt{TypeError}) across data splits, docstring conditions, and RAG settings. The distributions are used to show the overall error patterns. Then we go deeper into the analysis by comparing the distributions against one another to see if they differ with enough statistical significance.

In summary, by focusing on pass rates, paired comparisons, and error distributions, we provide a comprehensive and reproducible assessment of factors influencing functional correctness in real-world settings.

\section{Real-World Performance}
\label{sec:rq1}

This section addresses RQ1: \textit{How do Large Language Models (LLMs) perform on class-level code generation in real-world projects compared to synthetic benchmarks?} We evaluate all seven LLMs (Codestral, Deepseek-V3, GPT-4.1, GPT-5, GPT-OSS, Llama 4 Maverick, Qwen Coder 2.5) across three datasets: \textit{ClassEval} (100 synthetic classes), \textit{RealClassEval} pre-cutoff (200 classes from CodeSearchNet, likely seen during training), and \textit{RealClassEval} post-cutoff (200 classes from post-cutoff projects, unseen). The performance metric is the pass rate, defined as the proportion of passed test cases per class.

\subsection{Approach}
\label{subsec:rq1_approach}

To compare LLM performance across synthetic and real-world datasets, we employed a multi-step statistical approach combining per-class and pooled analyses.

\subsubsection{Data Processing}
We processed test reports for each LLM per dataset. Per-class pass rates were computed for analysis. For the pooled analysis, test cases were aggregated into total passed and failed counts per dataset and LLM.

\subsubsection{Kruskal-Wallis Test}
We used the Kruskal-Wallis test~\citep{kruskal1952use} to compare per-class pass rates across the three datasets (ClassEval, pre-cutoff, post-cutoff) for each LLM. In the current context, the null hypothesis states that the pass rate distributions are identical across datasets, while the alternative hypothesis states that at least one dataset's distribution differs.
\begin{align}
H_0: F_{ClassEval} = F_{pre-cutoff} = F_{post-cutoff} \label{eq:h0_kruskal}
\end{align}

\begin{align}
H_A: F_j \neq F_k & \text{ for at least one pair } (j, k), \notag \\
& \text{ where } j, k \in \{ \text{ClassEval}, \text{pre-cutoff}, \text{post-cutoff} \}, j \neq k \label{eq:h1_kruskal}
\end{align}
The Kruskal-Wallis test is a non-parametric alternative to one-way ANOVA, suitable for non-normal data and independent samples across three or more groups. The non-normality of pass rates was confirmed by the Shapiro-Wilk test~\citep{shapiro1965analysis}, justifying the non-parametric approach.

\subsubsection{Mann-Whitney U Tests}
For significant Kruskal-Wallis results, we conducted pairwise Mann-Whitney U tests~\citep{mann1947test} (ClassEval vs. pre-cutoff, ClassEval vs. post-cutoff, pre-cutoff vs. post-cutoff) to identify specific differences. The Mann-Whitney U test is a nonparametric test suitable for unpaired samples (i.e., distinct classes across datasets) and non-normal distributions. We applied Benjamini-Hochberg~\citep{benjamini1995controlling} False Discovery Rate (FDR) correction globally across all 21 tests (3 comparisons per LLM, 7 LLMs) to control for multiple comparisons. The pairwise tests pinpoint which datasets differ, while FDR correction balances Type I error across all tests.

\subsubsection{Effect Sizes and Confidence Intervals}
We computed Cliff's Delta~\citep{cliff1993dominance,macbeth2011cliff} to quantify the magnitude of pairwise differences, with thresholds~\citep{cohen2013statistical,rahman2024automatic}: negligible ($<0.147$), small ($<0.33$), medium $(<0.474$), large ($\geq0.474$). Bootstrap 95\% confidence intervals (CIs) were calculated for Cliff's Delta to quantify uncertainty, using 1,000 iterations. Cliff's Delta provides an effect size for ordinal data, and CIs ensure robust estimation of differences. Although Cliff's Delta help us get the effect size of differences, it does not quantify the actual differences. Hence, we also calculate the mean differences between the distributions with bootstrap 95\% CIs for an easy interpretation.

\subsubsection{Pooled Analysis}
We performed a chi-square analysis~\citep{pearson1900x,fisher1922interpretation} on pooled test case counts (passed vs. failed) using 2×2 contingency tables for pairwise dataset comparisons. The chi-square test is appropriate for categorical outcomes. Cramér's V~\citep{cramer1999mathematical} quantified effect sizes~\citep{cohen2013statistical}: negligible ($<0.1$), small ($<0.3$), medium ($<0.5$), large ($\geq0.5$), with bootstrap CIs for proportion differences. FDR correction was applied globally across all 21 tests. The pooled analysis provides an aggregate view of performance, complementing per-class results and ensuring robustness.

\subsection{Findings}
\label{subsec:rq1_findings}

\begin{figure}
    \centering
    \includegraphics[width=\columnwidth]{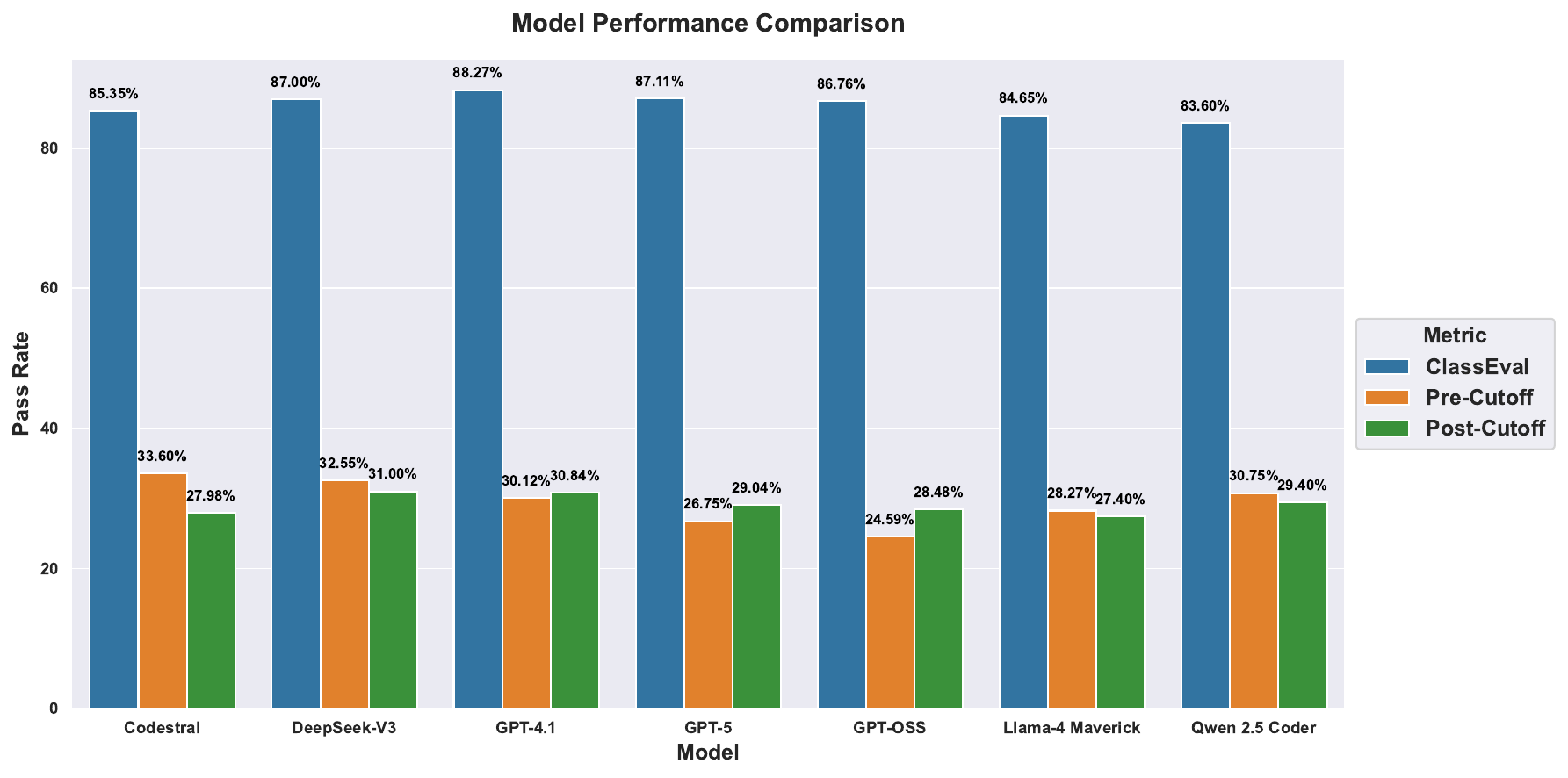}
    \caption{Performance of different LLMs on ClassEval, Pre-Cutoff, and Post-Cutoff data.}
    \label{fig:model_performance_across_datasets}
\end{figure}

The analysis revealed significant performance differences across datasets, with \textit{ClassEval} consistently outperforming both versions of \textit{RealClassEval}. Figure~\ref{fig:model_performance_across_datasets} shows the overall trend of performance of different LLMs across the three datasets under investigation, while Tables~\ref{tab:rq1_perclass} and~\ref{tab:rq1_pooled} summarize the per-class analysis result and pooled analysis result, respectively.

\subsubsection{Per-Class Analysis}
The Kruskal-Wallis test revealed significant differences in per-class pass rates across datasets for all LLMs (all $p < 1\times10^{-14}$). Pairwise Mann-Whitney U tests (Table~\ref{tab:rq1_perclass}) confirmed significant differences for ClassEval vs. pre-cutoff and ClassEval vs. post-cutoff, with mean pass rate differences of 46.93--61.24\% and 43.96--57.05\%, respectively, and medium to large effect sizes (Cliff's Delta between 0.46 and 0.64). pre-cutoff vs. post-cutoff comparisons showed no significant differences for six of seven models after FDR correction. GPT-OSS showed a statistically significant difference (FDR-adjusted $p = 0.019$), though the effect size remained negligible (Cliff's Delta $= -0.125$). All pre-cutoff vs. post-cutoff comparisons had negligible practical effect sizes. Power analysis across all seven models indicated high statistical power (0.974--0.983), ensuring reliable detection of differences.

\begin{table}[t]
\centering
\caption{Per-Class Pass Rate Analysis, condensed across all seven models (full per-model breakdown in Table~\ref{tab:rq1_perclass_full}, Appendix~\ref{app:full_tables}).}
\label{tab:rq1_perclass}
\resizebox{\textwidth}{!}{
\begin{tabular}{l|c|c|c}
\toprule
\textbf{Comparison} & \textbf{Mean Diff. Range (95\% CI)} & \textbf{Effect Size Range} & \textbf{\# Models Significant / 7} \\
\midrule
ClassEval vs. Pre-Cutoff & 46.93--61.24\% & Medium--Large & 7 / 7 \\
ClassEval vs. Post-Cutoff & 43.96--57.05\% & Medium--Large & 7 / 7 \\
Pre-Cutoff vs. Post-Cutoff & -9.05--4.57\% & Negligible & 1 / 7 (GPT-OSS) \\
\bottomrule
\end{tabular}}
\end{table}

\subsubsection{Pooled Analysis}
The chi-square analysis (Table~\ref{tab:rq1_pooled}) on pooled test case counts confirmed significant differences for ClassEval vs. pre-cutoff and ClassEval vs. post-cutoff, with pass rate differences of 51.75--62.17\% and 54.20--58.28\%, respectively, and medium to large effect sizes (Cramér's V ranging between 0.44 to 0.56). No significant differences were found for pre-cutoff vs. post-cutoff across all models (all FDR-adjusted $p > 0.07$, negligible effect sizes). Bootstrap 95\% CIs confirmed that the observed gaps are robust.

\begin{table}[t]
\centering
\caption{Pooled Chi-Square Analysis, condensed across all seven models (full per-model breakdown in Table~\ref{tab:rq1_pooled_full}, Appendix~\ref{app:full_tables}).}
\label{tab:rq1_pooled}
\resizebox{\textwidth}{!}{
\begin{tabular}{l|c|c|c}
\toprule
\textbf{Comparison} & \textbf{Mean Diff. Range (95\% CI)} & \textbf{Effect Size Range (Cram\'{e}r's V)} & \textbf{\# Models Significant / 7} \\
\midrule
ClassEval vs. Pre-Cutoff & 51.75--62.17\% & Medium--Large & 7 / 7 \\
ClassEval vs. Post-Cutoff & 54.20--58.28\% & Large & 7 / 7 \\
Pre-Cutoff vs. Post-Cutoff & -3.89--5.62\% & Negligible & 0 / 7 (all FDR-adj. $p > 0.10$) \\
\bottomrule
\end{tabular}}
\end{table}

\subsection{Interpretation}
\label{subsec:rq1_interpretation}

The findings indicate that LLMs perform significantly better on synthetic benchmarks (ClassEval, 83.60--88.27\%) than real-world projects (pre-cutoff and post-cutoff, 24.59--33.60\%), answering RQ1. pre-cutoff and post-cutoff classes showed no significant performance difference for six of seven models, with negligible effect sizes across all models. It suggests that we find no evidence that corpus-level familiarity with related code patterns meaningfully improves real-world class-level code generation, rather than a direct test of memorization per se. This distinction matters because CodeSearchNet is a function-level corpus with (comment, code) pairs, not a corpus of complete classes. Even where a model's training data overlapped with CodeSearchNet, it would have encountered the constituent functions individually rather than the class-level artifacts we evaluate here, and reconstructed as such in our dataset. Our seen versus unseen split should therefore be read as an approximation of corpus-level exposure to related code, not a confirmed measure of exposure to the specific evaluation artifacts, and we discuss this construct-validity limitation further in Section~\ref{sec:limitations}. Our results further show that the real-world performance range (24.59--33.60\%) is directionally consistent with the recent work called RepoClassBench~\citep{deshpande2024class}, which reports 24--34\% baseline performance on Python classes using natural language descriptions and repository context rather than just class skeletons. We note this alignment as suggestive rather than conclusive, since the two studies use non-comparable evaluation instruments (our PYNGUIN-generated automated tests versus RepoClassBench's repository-native test suites) and differ in task setup. We therefore do not treat this as a controlled comparison capable of supporting strong claims that specification richness generally overcomes implementation limitations.

The high pass rate of \textit{ClassEval} compared to \textit{RealClassEval} was significant, but the reasons were initially unclear, as the large effect sizes (Cliff's Delta 0.46--0.64) and pass rate differences did not fully explain the gap. To investigate, we conducted a mixed-effects model~\citep{laird1982random,carey2001mixed} analysis, including dataset, test count, and LLM-generated code metrics (lines of code, method count, cyclomatic complexity, and class coupling) as predictors, with LLM as a random effect. We chose these $4$ static metrics because they represent both the length of the code as well as the complexity of the code. The model showed significant dataset effects (e.g., $\beta = -1.060$, $p < 0.001$) and a negative test count effect ($\beta = -0.014$, $p < 0.001$). While lines of code ($\beta = -0.002$, $p = 0.006$) and cyclomatic complexity ($\beta = 0.043$, $p < 0.001$) showed statistical significance, their effects were small relative to the dataset effect, indicating that code complexity metrics explain only a minor portion of the performance gap.

To test whether the synthetic-versus-real-world gap reflects genuine implementation difficulty rather than dependency exposure, we split classes into context-free and dependency-heavy subgroups by class coupling and re-fit the mixed-effects model within each. The dataset effect persisted at comparable magnitude in both subgroups ($\beta = -0.90$ to $-0.86$ context-free; $\beta = -1.10$ to $-0.95$ dependency-heavy; all $p < 0.001$). This indicates the gap is not primarily an artifact of dependency exposure or class coupling. If anything, the gap is slightly larger among dependency-heavy classes, suggesting real-world implementation difficulty extends beyond what coupling alone would predict.

We further tested whether the gap is attributable to differential test-suite coverage by adding branch coverage as a covariate to the same model on classes with available per-class coverage data. The dataset effect was modestly attenuated but remained large and highly significant with coverage included ($\beta = -0.902$ vs.\ $-0.940$ for ClassEval--pre-cutoff; $\beta = -0.950$ vs.\ $-0.976$ for ClassEval--post-cutoff, with vs.\ without coverage; all $p < 0.001$), and coverage itself was a significant predictor ($\beta = 0.149$, $p = 0.001$). Together with the subgroup analysis above, this indicates the synthetic-versus-real-world performance gap is not primarily attributable to the confounds we can directly measure within our design: dependency exposure, class coupling, or test-suite coverage differences. This analysis cannot, however, rule out the confound shared by every class in our evaluation: the complete absence of repository-level context. The evidence therefore strongly supports that real-world, context-free class generation remains challenging for current LLMs, though it does not fully separate class-level implementation difficulty from the effects of missing project-specific context.

We then qualitatively analyzed test suites for randomly selected classes. The analysis revealed that \textit{ClassEval} test suites used simple equality assertions (e.g., numerical, dictionary checks) with no external dependencies, while \textit{RealClassEval} suites involved complex type metadata checks and system dependencies (e.g., \texttt{numpy}, \texttt{MCPContext}), leading to errors like \texttt{AttributeError}. This suggests that \textit{ClassEval}'s simpler test suites, despite higher test counts, enable higher pass rates by requiring straightforward functionality, unlike \textit{RealClassEval}'s complex, dependency-heavy tests.

\begin{tcolorbox}
    \textbf{RQ1 Findings:} LLMs achieve significantly higher pass rates on synthetic benchmarks compared to real-world projects. Six of seven models show no significant performance differences between pre-cutoff and post-cutoff real-world data, with all models exhibiting negligible effect sizes. We find no evidence that corpus-level exposure to related code patterns meaningfully improves real-world class-level performance, though we cannot rule out post-training exposure, distillation, or other contamination pathways for closed models, and note that CodeSearchNet's function-level granularity limits how directly this speaks to memorization of class-level artifacts specifically.
\end{tcolorbox}

\section{Impact of Docstrings}
\label{sec:rq2}

To understand the impact of documentation on code generation quality, we conducted a comprehensive ablation study examining how varying levels of docstring completeness affect the functional correctness of LLM-generated classes. This section addresses RQ2: \textit{Does the presence of docstrings improve the functional correctness of LLM-generated classes?} We evaluate seven LLMs (Codestral, Deepseek-V3, GPT-4.1, GPT-5, GPT-OSS, Llama 4 Maverick, Qwen Coder 2.5) across two versions of \textit{RealClassEval}: pre-cutoff and post-cutoff. Like before, the performance metric is the pass rate, defined as the proportion of passed test cases per class.

\subsection{Approach}
\label{subsec:rq2_approach}

To assess the effect of docstring completeness on LLM performance, we employed a within-subjects experimental design combining per-class and pooled analyses. The per-class analysis preserves granularity by comparing pass rates across docstring conditions, while the pooled analysis aggregates test case outcomes to assess overall trends. The following details the steps.

\subsubsection{The Ablation Study}

We evaluated all seven LLMs across three documentation conditions to isolate the impact of documentation completeness on class-level code generation. This systematic ablation study controls for snippet-specific characteristics by testing each class under all documentation conditions, enabling precise measurement of documentation effects.

Our experimental design implements three documentation levels that reflect realistic development scenarios: \textbf{(1) full docstrings} with complete documentation for all program units, including comprehensive class descriptions, method purposes, parameter specifications, and return value descriptions; \textbf{(2) partial docstrings} with documentation for a randomly selected subset of program units, simulating incomplete coverage common in real projects; and \textbf{(3) no docstrings} providing only function signatures and class structure.

The partial docstrings condition represents the most realistic scenario in software development, created by systematically removing docstrings from the full condition through random selection. We ensured that at least one documented component remained per class while varying coverage patterns. This process yielded partial documentation that retained an average of 60\% of the original content (mean: 127 tokens vs. 203 tokens for full condition), closely mimicking incomplete documentation found in production codebases.

To illustrate this methodology, we demonstrate the ablation process using the skeleton extracted from the {\tt SubdomainMiddleware} class shown in Listing~\ref{lst:example_skeleton}. The full docstrings condition includes complete documentation as extracted from the original implementation. From this baseline, our random removal process eliminates selected docstrings to create the partial condition, as shown in Listing~\ref{lst:partial_example}:

\begin{lstlisting}[
    language=Python,
    basicstyle=\ttfamily\small,
    breaklines=true,
    caption={Partial docstrings condition - random removal of some docstrings from {\tt SubdomainMiddleware}.},
    label=lst:partial_example,
    captionpos=t,
    commentstyle=\color{codegray},
    keywordstyle=\color{codeblue}\bfseries,
    stringstyle=\color{codegreen},
    showstringspaces=false,
    showspaces=false,
    numbers=none
]
class SubdomainMiddleware(object):
    def get_domain_for_request(self, request):
        pass

    def process_request(self, request):
        """
        Adds a "subdomain" attribute to the ``request'' parameter.
        """
        pass

\end{lstlisting}

In this example, the class-level docstring along with the first method docstring were removed while the second method docstring was retained, creating an incomplete specification that tests whether LLMs can infer the missing class purpose from available method documentation. The no docstrings condition removes all documentation, as shown in Listing~\ref{lst:no_example}:

\begin{lstlisting}[
    language=Python,
    basicstyle=\ttfamily\small,
    breaklines=true,
    caption={No docstrings condition - only structural information for {\tt SubdomainMiddleware}.},
    label=lst:no_example,
    captionpos=t,
    commentstyle=\color{codegray},
    keywordstyle=\color{codeblue}\bfseries,
    stringstyle=\color{codegreen},
    showstringspaces=false,
    showspaces=false,
    numbers=none
]
class SubdomainMiddleware(object):
    def get_domain_for_request(self, request):
        pass

    def process_request(self, request):
        pass

\end{lstlisting}

This systematic progression from complete to absent documentation enables precise measurement of how information availability affects generation quality. Each class in our dataset was evaluated under all three conditions, yielding paired samples that control for class-specific complexity while isolating documentation effects for repeated-measures statistical analysis.

\subsubsection{Friedman Test}

Given the continuous and bounded nature of our outcome variable (pass rate $\in [0, 1]$) and the paired structure of our data, we employed non-parametric statistical methods. The Friedman test~\citep{friedman1937use,friedman1940comparison} compared pass rates across three docstring conditions (full, partial, no) for each model-dataset combination, testing the null hypothesis that pass rate distributions are identical versus the alternative that at least one condition differs.

\subsubsection{Wilcoxon Signed Rank Test}
For significant Friedman results, post-hoc Wilcoxon signed-rank tests~\citep{wilcoxon1992individual} assessed pairwise differences (full vs. partial, full vs. no). High skewness ($|skewness| > 1$ in 11 of 28 comparisons) prompted sign tests to validate Wilcoxon results.

Similar to RQ1, other steps of the analysis include a pooled chi-square analysis on aggregated pass-fail counts, using 2×2 contingency tables for pairwise comparisons. Benjamini-Hochberg False Discovery Rate (FDR) correction ($\alpha = 0.05$) was applied separately per dataset (7 Friedman, 14 Wilcoxon, 14 chi-square tests per dataset), treating pre-cutoff and post-cutoff as independent experiments on different code populations. Effect sizes included Cliff's Delta (negligible $< 0.147$, small $< 0.33$, medium $< 0.474$, large $\geq0.474$) and Cramér's V (negligible $< 0.1$, small $< 0.3$, medium $< 0.5$, large $\geq0.5$), with bootstrap 95\% confidence intervals (CIs).

\subsubsection{Temporal Analysis}
A temporal analysis computed benefit ratios (\# improved / \# worsened) and net benefits (\# improved / \# unchanged × 100) for pre-cutoff and post-cutoff datasets, with a Wilcoxon test comparing model-wise ratios to assess whether docstring benefits differ between seen and unseen data.

\subsection{Findings}
\label{subsec:rq2_findings}

\begin{figure}[h]
    \centering
    \includegraphics[width=\columnwidth]{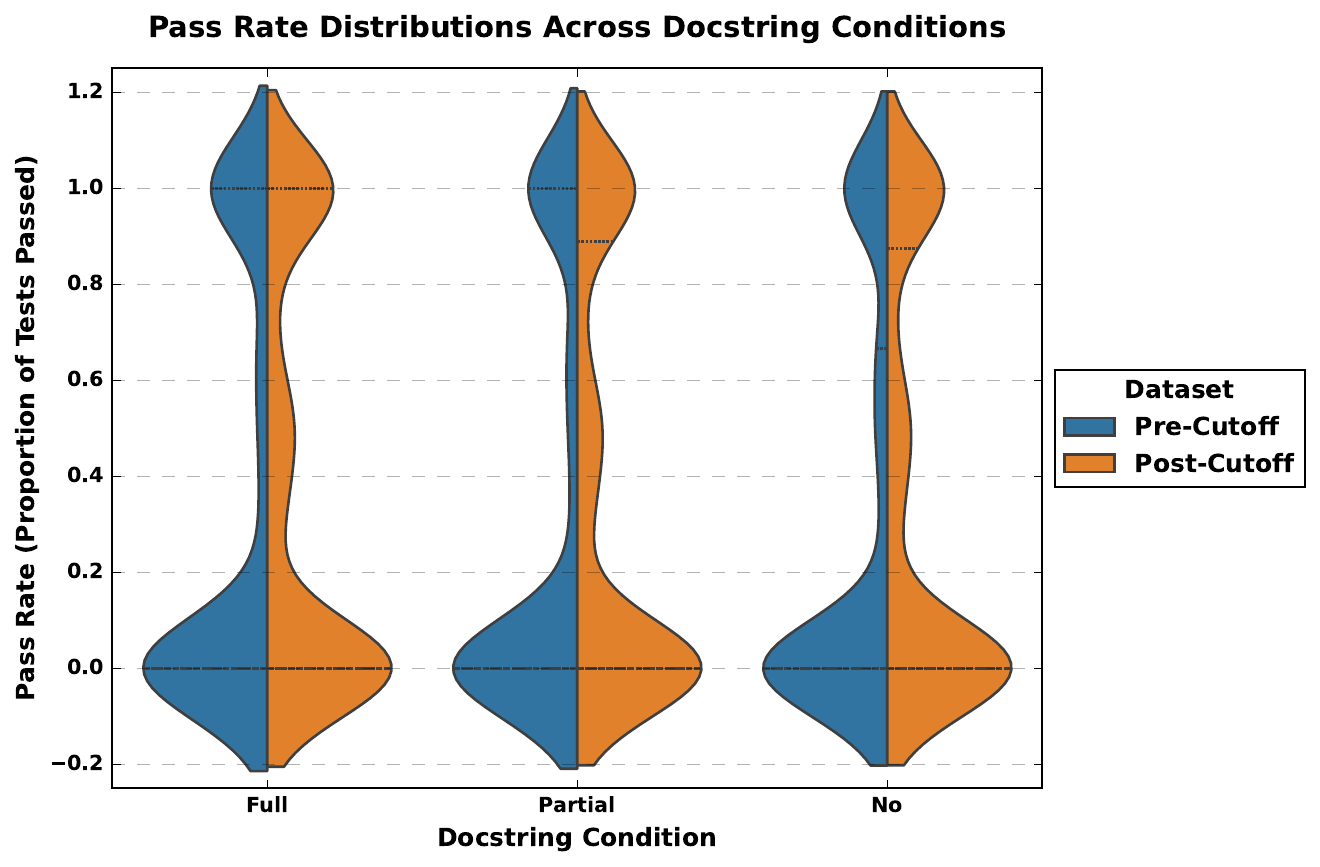}
    \caption{Violin plot of per-class pass rate distributions across docstring conditions (full, partial, no) for Pre-Cutoff and Post-Cutoff datasets, aggregated across all models. Violin plots use kernel density estimation to show data distribution; the y-axis extends beyond [0,1] due to KDE smoothing~\citep{silverman2018density}, though all actual pass rate values are bounded between 0 and 1.}
    \label{fig:pass_rate_violins}
\end{figure}

The analysis revealed limited and model-specific effects of docstrings on code generation quality. Figure~\ref{fig:pass_rate_violins} shows pass rate distributions across docstring conditions, with only subtle differences between conditions. Tables~\ref{tab:summary_results}, \ref{tab:model_results}, and~\ref{tab:temporal_analysis} summarize per-class, model-specific, and temporal results, respectively.

\subsubsection{Per-Class Analysis}

The Friedman test revealed significant differences in pass rates across docstring conditions for three model-dataset combinations after FDR correction: Codestral (pre-cutoff: $p=1.02e-03$, $p\textsubscript{FDR}=0.007$), GPT-4.1 (pre-cutoff: $p=2.93e-03$, $p\textsubscript{FDR}=0.010$), and Deepseek-V3 (post-cutoff: $p=6.81e-03$, $p\textsubscript{FDR}=0.048$). However, pairwise Wilcoxon tests (Table~\ref{tab:summary_results}) showed only two significant improvements after FDR correction: Codestral (pre-cutoff, full vs. partial: $p\textsubscript{FDR}=0.049$) and Deepseek-V3 (post-cutoff, full vs. partial: $p\textsubscript{FDR}=0.048$), with mean differences of -0.058 (-2.72\%) and -0.047 (-1.51\%) tests, respectively. The negative values indicate that full docstrings outperform partial docstrings. Notably, the full vs. no comparison yielded no significant results after FDR correction.

All Cliff's Delta values ranged from 0.003 to 0.077 (negligible effect sizes), reflecting minimal practical magnitudes. High skewness prompted sign tests for 11 comparisons, with two showing significance after FDR correction, corroborating the corresponding Wilcoxon results. The remaining 26 of 28 Wilcoxon comparisons showed no significant effects, indicating that for most models, docstrings do not significantly impact functional correctness.

\begin{table}[t]
    \centering
    \caption{Summary of docstring ablation results across all models and datasets with respect to the number of test cases. Negative differences indicate full docstrings cause more test cases to pass compared to the alternative condition.}
    \label{tab:summary_results}
    \resizebox{\columnwidth}{!}{
    \begin{tabular}{l|c|c|c|c|c|c}
    \toprule
    \textbf{Comparison} & \textbf{Mean \% Diff Range} & \textbf{Improved} & \textbf{Worsened} & \textbf{Unchanged} & \textbf{\makecell{Significant\\(after FDR Correction)}} & \textbf{Effect Size} \\
    \midrule
    Full vs. Partial & -0.15\% to -2.72\% & 97 & 147 & 2,371 & 2/28 & Negligible \\
    \midrule
    Full vs. No & -0.44\% to -3.13\% & 113 & 235 & 2,267 & 0/28 & Negligible \\
    \midrule \midrule
    \textbf{Combined} & \textbf{-0.15\% to -3.13\%} & \textbf{210} & \textbf{382} & \textbf{4,638} & \textbf{2/56} & \textbf{Negligible} \\
    \bottomrule
    \end{tabular}}
\end{table}

\subsubsection{Pooled Analysis}

The chi-square analysis on pooled test case counts showed no significant differences after FDR correction for either dataset (all p\textsubscript{FDR} $> 0.05$). Cramér's V values ranged from 0.002 to 0.066 (negligible effect sizes), confirming that aggregate test-level effects are minimal. This finding suggests that while individual snippets may show improvements, these effects do not accumulate to produce population-level differences in test case outcomes.

\subsubsection{Model-Specific Analysis}

Table~\ref{tab:model_results} presents model-specific improvements with full docstrings. Only two of 14 model-dataset combinations showed statistically significant benefits: Codestral in pre-cutoff data (full vs. partial: 2.72\%, p\textsubscript{FDR}=0.049) and Deepseek-V3 in post-cutoff data (full vs. partial: 1.51\%, p\textsubscript{FDR}=0.048). The remaining 12 combinations showed no significant effects despite observed numerical differences ranging from 0.15\% to 3.13\%.

\begin{table}[t]
    \centering
    \caption{Model-specific performance improvements with full docstrings. Percentages represent pass rate increases when using full documentation relative to partial or no docstrings. \textbf{Bold} entries achieved statistical significance after FDR correction (p\textsubscript{FDR} $< 0.05)$.}
    \label{tab:model_results}
    \resizebox{\textwidth}{!}{
    \begin{tabular}{c|c|c|c|c}
    \toprule
    \textbf{Model} & \textbf{Dataset} & \textbf{Full vs. Partial (\%)} & \textbf{Full vs. No (\%)} & \textbf{Average Benefit (\%)} \\
    \midrule
    \multirow{2}{*}{Codestral} & Pre-Cutoff & \textbf{2.72} & 3.13 & 2.93 \\
     & Post-Cutoff & 1.01 & 0.80 & 0.90 \\
    \hline
    \multirow{2}{*}{Deepseek-V3} & Pre-Cutoff & 0.93 & 1.59 & 1.26 \\
     & Post-Cutoff & \textbf{1.51} & 1.32 & 1.42 \\
    \hline
    \multirow{2}{*}{GPT-4.1} & Pre-Cutoff & 0.51 & 2.11 & 1.31 \\
     & Post-Cutoff & 1.02 & 0.90 & 0.96 \\
    \hline
    \multirow{2}{*}{GPT-5} & Pre-Cutoff & 0.71 & 1.59 & 1.15 \\
     & Post-Cutoff & 0.15 & 0.93 & 0.54 \\
    \hline
    \multirow{2}{*}{GPT-OSS} & Pre-Cutoff & 1.33 & 1.58 & 1.45 \\
     & Post-Cutoff & 0.32 & 0.63 & 0.48 \\
    \hline
    \multirow{2}{*}{Llama 4 Maverick} & Pre-Cutoff & 1.88 & 2.58 & 2.23 \\
     & Post-Cutoff & 0.64 & 0.44 & 0.54 \\
    \hline
    \multirow{2}{*}{Qwen Coder 2.5} & Pre-Cutoff & 0.41 & 1.91 & 1.16 \\
     & Post-Cutoff & 0.93 & 1.36 & 1.15 \\
    \bottomrule
    \end{tabular}}
\end{table}
\subsubsection{Temporal Analysis}

Table~\ref{tab:temporal_analysis} presents the comparative impact of docstrings across temporal boundaries. The benefit ratio (\# improved / \# worsened) was 1.58:1 for pre-cutoff data and 2.14:1 for post-cutoff data, representing a numerical difference of 35\%. However, a Wilcoxon test comparing model-wise benefit ratios was non-significant (p=0.326), indicating no statistically reliable difference in docstring effectiveness between pre-cutoff and post-cutoff data. This null result reflects substantial model-specific variability: Codestral showed stronger post-cutoff benefits (benefit ratio 0.61 post vs. 0.30 pre for full vs. no), while Deepseek-V3 showed the opposite pattern (0.37 post vs. 0.60 pre). The lack of a consistent temporal effect suggests that docstring effectiveness depends primarily on model-specific factors rather than whether code was in the training data.

\begin{table}[t]
    \centering
    \caption{Temporal analysis of docstring impact. Benefit ratio is the ratio of classes that improved to classes that worsened with full docstrings. The Wilcoxon test comparing Pre-Cutoff vs. Post-Cutoff benefit ratios across models was non-significant.}
    \label{tab:temporal_analysis}
    \resizebox{\textwidth}{!}{
    \begin{tabular}{l|c|c|c|c}
    \toprule
    \textbf{Dataset} & \textbf{Classes Helped} & \textbf{Classes Hurt} & \textbf{Benefit Ratio} & \textbf{Net Benefit (\%)} \\
    \midrule
    Pre-Cutoff & 210 (8.9\%) & 133 (5.6\%) & 1.58:1 & 3.3\% \\
    \midrule
    Post-Cutoff & 172 (7.1\%) & 80 (3.3\%) & 2.14:1 & 3.8\% \\
    \bottomrule
    \end{tabular}}
\end{table}

\subsection{Interpretation}
\label{subsec:rq2_interpretation}

The findings provide limited evidence that docstrings improve LLM code generation quality, addressing RQ2 with a nuanced answer: \textbf{for most models, docstrings do not significantly improve functional correctness}. Only 2 of 7 models (Codestral and Deepseek-V3) showed statistically significant benefits in specific conditions, and even these effects were negligible in magnitude. This pattern, combined with adequate statistical power (0.78–0.87) obtained from power analysis, suggests that docstring effects are genuinely minimal rather than obscured by insufficient sample sizes.

\subsubsection{Interpreting Limited Effects}

The lack of widespread significance after FDR correction indicates that raw numerical improvements (averaging 1.25\% across all comparisons) are largely attributable to random variation and multiple testing rather than true effects. The fact that pooled chi-square tests found zero significant results reinforces this interpretation: individual snippet-level fluctuations do not aggregate into reliable population-level patterns.

The high skewness in 11 of 28 comparisons reveals that most classes show zero difference across docstring conditions, with occasional large outliers driving mean differences. This suggests docstring benefits may be highly context-dependent, helping only in specific scenarios rather than providing universal improvements. The negligible Cliff's Delta values (all $< 0.15$) confirm minimal practical effect sizes, even for the two statistically significant findings.

\subsubsection{Model-Specific Heterogeneity}

The restriction of significant effects to two models (Codestral and Deepseek-V3) suggests that docstring processing may depend on model architecture or training procedures. Possible explanations include:

\begin{enumerate}[leftmargin=*]
    \item \textbf{Training data composition}: Models trained on repositories with comprehensive docstring conventions may learn stronger associations between documentation and correct implementations.
    \item \textbf{Attention mechanisms}: Architectural differences in how models weight docstring tokens versus signature tokens could affect documentation utilization.
    \item \textbf{Model capacity}: Smaller or more constrained models may rely more heavily on explicit documentation, while larger models can infer intended behaviour from signatures alone.
\end{enumerate}

The fact that five of seven models showed no significant effects despite testing on the same code snippets indicates that documentation alone is insufficient for most current LLMs. This contrasts with conventional software engineering wisdom emphasizing documentation importance~\citep{boehm1984software,sommerville2011software}, suggesting that LLMs may process code differently than human developers.

Future research should investigate: (1) whether specific docstring formats (e.g., type hints vs. natural language descriptions) provide differential benefits; (2) whether docstrings interact with other contextual factors like code complexity; (3) why certain models (Codestral, Deepseek-V3) show sensitivity while others do not; and (4) whether docstring benefits emerge for more complex classes beyond our dataset's scope.

\begin{tcolorbox}
    \textbf{RQ2 Findings:} Comprehensive statistical analysis with appropriate multiple comparison corrections found significant improvements in only 2 of 7 models, and only for the full vs. partial comparison. Effect sizes were uniformly negligible.
\end{tcolorbox}

\section{Impact of Retrieval-Augmented Generation (RAG)}
\label{sec:rq3}

This section investigates whether retrieval-augmented generation (RAG) can further enhance functional correctness for unseen code by providing relevant examples from seen data. RQ3: \textit{Can retrieval-augmented generation (RAG) using seen data improve LLM performance on unseen class-level code generation tasks?} We evaluate the same seven LLMs (Codestral, Deepseek-V3, GPT-4.1, GPT-5, GPT-OSS, Llama 4 Maverick, Qwen Coder 2.5) on the \textit{RealClassEval} post-cutoff dataset with and without RAG-retrieved examples. The examples were retrieved from \textit{RealClassEval} pre-cutoff data. The same performance metric of pass rate was used.

We emphasize a framing distinction before presenting results: RAG is an external retrieval-and-prompting strategy layered on top of a model, not a property of the model itself. The findings in this section therefore speak to the performance of a specific \textit{retrieval-augmented pipeline} under our experimental conditions, and should not be conflated with claims about base LLM capability made in RQ1 and RQ2, which evaluate models without any retrieved context.

\subsection{Approach}
\label{subsec:rq3_approach}

To assess RAG's impact on unseen code generation, we employed a controlled experimental design comparing LLM performance with and without retrieved documentation from seen data.

\subsubsection{Data Preparation}

The evaluation utilized \textit{RealClassEval}'s post-cutoff dataset, ensuring all test cases were from projects created after model training cutoffs. For each target class skeleton, RAG retrieved the top $k=2$ most similar class skeletons from pre-cutoff data using cosine similarity on embeddings obtained from \textbf{GraphCodeBert}~\citep{guo2020graphcodebert}, then appended their complete implementations (signature, docstrings, and body) to the prompt.

We selected $k=2$ to balance retrieval effectiveness with computational feasibility across our large-scale evaluation. This choice enables systematic investigation of RAG's impact on class-level generation while maintaining practical scalability across seven models and 400 classes. Real-world class implementations vary significantly in length, and larger values of $k$ would create prohibitively long prompts for many classes, potentially exceeding model context windows or incurring high computational costs that would limit experimental scope. Yang et al.~\citep{yang2025empirical} also frame the trade-off between retrieval-augmented performance gains and computational cost as a recognized, first-order design consideration in retrieval-augmented code generation pipelines in general. Our focus centers on demonstrating \textit{whether} retrieval-augmented generation improves class-level code generation rather than optimizing retrieval hyperparameters, which is an important but distinct research question that would require dedicated parameter space exploration. The $k=2$ configuration provides sufficient retrieved context to demonstrate RAG mechanisms (showing concrete implementation patterns and examples) while ensuring consistent evaluation conditions across all experimental settings. This choice is also consistent with prior findings in retrieval-augmented code generation. Chen et al.~\citep{chen2026repository} report that in-context learning performance for repository-level code generation peaks with exactly two retrieved/demonstration examples across multiple models, with additional examples degrading performance due to context saturation. More broadly, retrieval-augmented approaches to code-related tasks do not consistently benefit from larger $k$: Meng et al.~\citep{meng2025more} find that top-1 retrieval outperforms larger retrieval sets in code review generation, with additional retrieved items introducing redundancy and conflicting cues, and Wang et al.~\citep{wang2023codet5plus} use $k=1$ in the retrieval-augmented generative setting for CodeT5+. Shi et al.~\citep{shi2022race} tested $k \in \{1,3,5,7,9\}$ for retrieval-augmented commit message generation and found performance was generally stable across this range, suggesting diminishing returns to exhaustive $k$-tuning once $k$ is small. Together, this literature supports treating $k=2$ as a reasonable, literature-consistent design choice rather than an arbitrary one. It does not, however, substitute for an ablation on our own data, which we discuss as a direction for future work given the computational cost constraints already noted above.

\subsubsection{Study Design}

Like RQ2, three docstring conditions were tested to investigate RAG's interaction with documentation completeness: (1) \textbf{full docstrings}; (2) \textbf{partial docstrings}; and (3) \textbf{no docstrings}. Each condition was evaluated both with RAG (retrieved examples included) and without RAG, yielding paired samples for within-subjects analysis.

\subsubsection{Statistical Analysis}

No new statistical methodology has been used in RQ3 that has not been used in the previous two RQs. All the statistical tests applied in this study have been discussed in the previous section(s). Here is a quick summary of what steps we have taken to answer RQ3. 

Given the paired structure (each snippet tested with and without RAG) and non-normal pass rate distributions, we again applied the Wilcoxon signed-rank test to compare pass rate differences (RAG - non-RAG) for each model-docstring combination. To control Type I errors while preserving statistical power, we applied the Benjamini-Hochberg False Discovery Rate (FDR) correction ($\alpha = 0.05$) \textit{separately for each docstring condition} (7 model comparisons per condition). The decision to apply FDR correction separately per docstring type (rather than globally across all 21 tests) warrants justification. We treat the three docstring conditions as substantively distinct experimental contexts: full docstrings provide complete specifications (testing RAG's value when information is abundant), partial docstrings create information gaps (testing RAG's gap-filling capability), and no docstrings represent minimal guidance (testing RAG's effectiveness without structural anchors). These conditions address fundamentally different research sub-questions about RAG's mechanism, making separate FDR families appropriate. This choice was made because we want to understand RAG's effect on different docstring types individually, and not collectively.

FDR-adjusted p-values are reported, with statistical significance defined as $p\textsubscript{FDR} < 0.05$. Cliff's Delta measured effect size (negligible $< 0.147$, small $< 0.33$, medium $< 0.474$, large $\geq0.474$) with 95\% bootstrap confidence intervals. For distributions with high skewness ($|skewness| > 1$), sign tests validated Wilcoxon results. Power analysis for paired designs estimated the study's ability to detect observed effect sizes.

\subsection{Findings}
\label{subsec:rq3_findings}

RAG demonstrates significant benefits for code generation with partial docstrings, with 5 of 7 models showing statistically significant improvements after FDR correction. Benefits are negligible with full docstrings and marginal with no docstrings, revealing a critical interaction between RAG and documentation completeness. Table~\ref{tab:rq3_summary} summarizes results across docstring conditions, while Table~\ref{tab:rq3_results} presents detailed model-specific findings organized by documentation level.

\begin{table}[t]
    \centering
    \caption{Summary of RAG effects across docstring conditions on Post-Cutoff dataset. Positive mean differences indicate RAG outperforms non-RAG. Per-type FDR correction applied separately to each condition (7 tests per condition).}
    \label{tab:rq3_summary}
    \begin{tabular}{l|c|c|c|c}
    \toprule
    \makecell{Docstring\\Condition} & \makecell{Mean Percentage\\Point Improvement} & Improved & Worsened & \makecell{Significant\\(p\textsubscript{FDR}<0.05)} \\
    \midrule
    Full & 1.55 & 81 & 51 & 0/7 \\
    Partial & 4.25 & 125 & 55 & 5/7 \\
    No & 3.58 & 141 & 77 & 0/7 \\
    \midrule
    \textbf{Overall} & \textbf{3.11} & \textbf{347} & \textbf{183} & \textbf{5/21} \\
    \bottomrule
    \end{tabular}
\end{table}

\subsubsection{RAG with Full Docstrings: No Significant Benefits}

When complete documentation is provided, RAG shows no significant improvements across all seven models (Table~\ref{tab:rq3_results}). Mean improvements range from -0.8\% (GPT-5) to 3.1\% (Deepseek-V3), with all $p\textsubscript{FDR} > 0.21$. This pattern suggests that when models receive comprehensive docstrings, retrieved examples provide minimal additional value.

\subsubsection{RAG with Partial Docstrings: Strong Significant Benefits}

RAG demonstrates its strongest and most consistent benefits when documentation is incomplete, with 5 of 7 models achieving statistical significance after FDR correction (Table~\ref{tab:rq3_results}). Significant improvements range from 4.2\% (Llama 4 Maverick, p\textsubscript{FDR}=0.024) to 6.9\% (Deepseek-V3, p\textsubscript{FDR}=0.023), with adequate statistical power (0.634–0.838) supporting these findings. Codestral (5.3\%, p\textsubscript{FDR}=0.030), GPT-4.1 (4.6\%, p\textsubscript{FDR}=0.023), and Qwen Coder 2.5 (5.2\%, $p\textsubscript{FDR}=0.024$) also show significant benefits. Only GPT-5 shows negative effects (-0.6\%, p\textsubscript{FDR}=0.739), suggesting model-specific architectural differences in utilizing retrieved context.

All Cliff's Delta values, however, remain negligible (0.049–0.078), indicating modest practical effect sizes despite statistical significance. However, power analysis confirms adequate detection capability for the observed effects (power 0.634–0.838 for significant results), validating that these are genuine improvements rather than false positives.

\subsubsection{RAG with No Docstrings: Marginal Non-Significant Benefits}

Without documentation, RAG shows positive trends for most models but fails to achieve statistical significance after FDR correction. Deepseek-V3 (5.8\%, p\textsubscript{raw}=0.016, p\textsubscript{FDR}=0.057) and Qwen Coder 2.5 (5.9\%, p\textsubscript{raw}=0.014, p\textsubscript{FDR}=0.057) approach significance, while Codestral shows a strong trend (5.7\%, p\textsubscript{raw}=0.061, p\textsubscript{FDR}=0.142). Power analysis reveals moderate detection capability (0.550–0.725), suggesting these effects may be genuine but require larger samples for definitive confirmation.

\begin{table}[t]
    \centering
    \caption{Model-specific RAG effects organized by docstring condition. A positive difference means RAG performed better, and vice versa. \textbf{Bold} entries indicate statistical significance.}
    \label{tab:rq3_results}
    \begin{tabular}{l|c|c|c|c}
    \toprule
    \textbf{Docstring} & \textbf{Model} & \textbf{Mean Pass} & \textbf{FDR-Adjusted $p$} & \textbf{Improved/} \\
    \textbf{Type} & & \textbf{Rate \% Diff.} & & \textbf{Worsened} \\
    \midrule
    \multirow{7}{*}{\textbf{Full}} 
    & Codestral & 1.79 & 0.536 & 11/6 \\
    & Deepseek-V3 & 3.07 & 0.536 & 16/9 \\
    & GPT-4.1 & 0.84 & 0.617 & 6/4 \\
    & GPT-5 & -0.77 & 0.798 & 12/13 \\
    & GPT-OSS & 2.37 & 0.536 & 17/8 \\
    & Llama 4 Maverick & 2.33 & 0.536 & 13/6 \\
    & Qwen Coder 2.5 & 0.94 & 0.617 & 6/5 \\
    \midrule
    \multirow{7}{*}{\textbf{Partial}} 
    & Codestral & 5.27 & \textbf{0.030} & 18/6 \\
    & Deepseek-V3 & 6.94 & \textbf{0.023} & 22/7 \\
    & GPT-4.1 & 4.63 & \textbf{0.023} & 16/5 \\
    & GPT-5 & -0.58 & 0.739 & 15/18 \\
    & GPT-OSS & 3.26 & 0.212 & 21/11 \\
    & Llama 4 Maverick & 4.25 & \textbf{0.024} & 18/4 \\
    & Qwen Coder 2.5 & 5.18 & \textbf{0.024} & 15/4 \\
    \midrule
    \multirow{7}{*}{\textbf{No}} 
    & Codestral & 5.74 & 0.142 & 22/8 \\
    & Deepseek-V3 & 5.84 & 0.057 & 23/11 \\
    & GPT-4.1 & 2.75 & 0.357 & 18/13 \\
    & GPT-5 & 1.07 & 0.667 & 20/18 \\
    & GPT-OSS & 2.20 & 0.488 & 18/12 \\
    & Llama 4 Maverick & 2.15 & 0.357 & 18/8 \\
    & Qwen Coder 2.5 & 5.93 & 0.057 & 22/7 \\
    \bottomrule
    \end{tabular}
\end{table}

\subsection{Interpretation}
\label{subsec:rq3_interpretation}

The findings reveal a critical interaction between RAG and documentation completeness, answering RQ3: the retrieval-augmented pipeline evaluated here significantly improves code generation quality for unseen tasks, but only when documentation is incomplete. This ``information gap hypothesis"~\citep{loewenstein1994psychology,golman2018information} suggests RAG's primary value lies in compensating for missing context rather than enhancing already-comprehensive specifications.

\subsubsection{The Information Gap Hypothesis}

The striking pattern--5/7 models significant with partial docstrings, 0/7 with full docstrings, 0/7 with no docstrings--supports a specific mechanistic interpretation. When documentation is complete, models already possess sufficient behavioural specifications; retrieved examples offer redundant information. When documentation is absent, models lack structured guidance to effectively leverage retrieved examples--the context is too noisy without docstrings to anchor understanding. The ``sweet spot" occurs with partial docstrings: enough structure to interpret retrieved examples, but insufficient detail to implement correctly without them.

\subsubsection{Practical Significance Despite Negligible Effect Sizes}

Although all Cliff's Delta values are negligible ($<0.147$), the improvements have practical importance. A 4–7\% pass rate increase translates to 0.12–0.28 additional test cases passed per class with a very restrictive assumption of mean test counts of 3–4 per class. For large codebases generating thousands of classes, this compounds to substantial quality gains. Moreover, combining RQ2's docstring benefits (mean 1.25\%) with RQ3's RAG benefits (4.25\% for partial docstrings) yields a potential 5.5\% cumulative improvement--equivalent to approximately 0.19 tests per class, or roughly 190 additional correct implementations per 1,000 generated classes. Additionally, RQ2 showed benefits of up to 2.7\% with full docstrings, while RQ3 shows 4.3–6.9\% with partial docstrings plus RAG. This suggests RAG and documentation are complementary strategies, with RAG providing greater relative benefits when documentation is lacking.

\subsubsection{Model-Specific Patterns and GPT-5's Anomaly}

Five models (Codestral, Deepseek-V3, GPT-4.1, Llama 4 Maverick, Qwen Coder 2.5) show consistent positive responses to RAG with partial docstrings, while GPT-5 exhibits negative effects (-0.8\% full, -0.6\% partial). This suggests architectural differences in context utilization. GPT-5 may suffer from ``context interference" (retrieved examples conflicting with its training-based priors) or may weight retrieved context differently than documentation. GPT-OSS shows positive trends (3.3\% partial) but insufficient power for significance, suggesting moderate responsiveness.

The consistency of the pattern across five diverse architectures (transformer variants, different training datasets, varying parameter counts) strengthens the generalizability of the ``information gap" hypothesis. RAG's effectiveness is not model-specific but rather a property of the interaction between retrieval quality and documentation completeness. However, we do acknowledge that further investigation is required to conclusively state why these inter-model differences are being observed.

\subsubsection{Interpret with Caution}

It is worth mentioning that power analysis revealed that this study's power varies substantially across conditions (0.059–0.838), with full and no docstring conditions showing inadequate power (mostly $<0.30$) to detect true effects. This explains why full docstrings show no significant results despite positive trends, meaning that the study cannot reliably distinguish small RAG benefits from noise. However, the partial docstring condition achieves adequate power (0.634–0.838) for significant results, validating these findings. Future work should increase sample sizes to achieve higher power for detecting true effects across all conditions.

\begin{tcolorbox}
    \textbf{RQ3 Findings:} This study provides systematic evidence that the effectiveness of the retrieval pipeline evaluated here depends critically on documentation completeness, with the strongest benefits occurring when models receive partial documentation. This finding bridges RQ2's insights on documentation effects and suggests a path toward more efficient code generation, though it speaks to this specific retrieval configuration rather than retrieval augmentation in general.
\end{tcolorbox}

\section{Error Analysis}
\label{sec:rq4}

To understand the nature of failures in LLM-generated code, this section addresses RQ4: \textit{What are the most common errors in LLM-generated class-level code?} We analyze error distributions across all experimental conditions from RQ1--RQ3, examining \textit{all failed test cases} across seven LLMs, three datasets, and multiple documentation and retrieval configurations. We characterize overall error patterns, compare synthetic versus real-world failure modes, and investigate how docstrings and RAG affect error types.

\subsection{Approach}
\label{subsec:rq4_approach}

To comprehensively characterize error patterns in LLM-generated code, we employed a multi-level statistical analysis examining error type distributions across experimental conditions.

\subsubsection{Error Classification and Data Processing}

We extracted error types from test execution reports across all conditions, parsing exception types from failure messages. Errors were classified into Python's standard exception taxonomy: \texttt{AttributeError}, \texttt{TypeError}, \texttt{AssertionError}, \texttt{ValueError}, \texttt{NameError}, \texttt{KeyError}, \texttt{IndexError}, \texttt{ZeroDivisionError}, \texttt{FileNotFoundError}, \texttt{ImportError}, \texttt{RuntimeError}, and \texttt{SyntaxError}. Non-standard errors, such as API specific exceptions, were aggregated into an ``Other'' category. For each failed class, we tabulated error counts by type, aggregating across test cases to create error frequency distributions.

\subsubsection{Statistical Analysis}

We applied chi-square tests of independence~\citep{pearson1900x,fisher1970statistical} to compare error type distributions across conditions, appropriate for categorical error frequency data. The analysis comprised four complementary comparisons:

\textbf{Overall Distribution Analysis}: Chi-square goodness-of-fit test~\citep{conover1999practical} against uniform distribution was applied to confirm non-uniform error patterns in order to establish that error types are non-randomly distributed across failures. The larger the chi-squared value, the greater the deviation from the uniform distribution, implying that there is a meaningful pattern worth investigating.

In our context, this test can be realized as follows: Let $p_i$ denote the probability of observing error type $i$ where $i \in \{\text{AttributeError}, \text{TypeError}, \ldots, \text{Other}\}$. We tested:

\begin{equation}
\begin{split}
H_0: p_i = \frac{1}{13} \quad \forall i \in \{1, 2, \ldots, 13\} \\
H_A: \exists i \text{ such that } p_i \neq \frac{1}{13}
\end{split}
\end{equation}

with test statistic $\chi^2 = \sum_{i=1}^{13} \frac{(O_i - E_i)^2}{E_i}$ where $O_i$ is the observed count and $E_i = \frac{\sum_{i=1}^{13} O_i}{13}$ is the expected count under uniform distribution. The test has $df = 12$ degrees of freedom.

Like previous RQs, Cramér's V for effect size and bootstrap 95\% confidence intervals (CIs) to quantify heterogeneity were also applied.

\textbf{Dataset-Wise Comparison}: Pairwise chi-square tests were applied to compare error distributions across ClassEval, pre-cutoff, and post-cutoff datasets for each model (3 comparisons × 7 models = 21 tests). Global Benjamini-Hochberg FDR correction ($\alpha = 0.05$) controlled Type I error across all tests, consistent with RQ1 methodology. Cramér's V quantified effect sizes with thresholds: negligible ($<0.1$), small ($<0.3$), medium ($<0.5$), large ($\geq$0.5).

\textbf{Docstring Impact}: Chi-square tests were again applied to compare error distributions for full vs. partial and full vs. no docstrings within pre-cutoff and post-cutoff datasets separately (2 comparisons per model-dataset pair). Per-dataset FDR correction (14 tests per dataset) followed RQ2's approach of treating temporal contexts as distinct experimental families.

\textbf{RAG Impact}: Another set of Chi-square tests was applied to compare RAG vs. non-RAG error distributions within each docstring condition for post-cutoff data (7 models per condition). Per-docstring-condition FDR correction (7 tests per condition) mirrored RQ3's methodology, treating documentation levels as substantively different information environments.

For all comparisons, we computed statistical power to detect observed effect sizes, ensuring reliable inference. Bootstrap CIs (1,000 iterations) provided robust uncertainty estimates for effect sizes. To contextualize statistical findings, we computed percentage point differences in error type frequencies, identifying which errors increase or decrease across conditions.

\subsection{Findings}
\label{subsec:rq4_findings}

The analysis revealed three major patterns: (1) a consistent error signature dominated by \texttt{AttributeError}, \texttt{TypeError}, and \texttt{AssertionError}; (2) fundamental differences in error types between synthetic benchmarks and real-world code; and (3) RAG-induced error substitution where retrieval prevents certain failures while introducing others. Table~\ref{tab:rq4_overall} summarizes the overall distribution, Tables~\ref{tab:rq4_dataset} and~\ref{tab:rq4_dataset_breakdown} detail dataset comparisons, and Table~\ref{tab:rq4_rag_substitution} characterizes RAG's error substitution mechanism.

\subsubsection{Overall Error Distribution}

Across all failed test cases, three error types dominate LLM-generated code failures (Table~\ref{tab:rq4_overall}). \texttt{AttributeError} accounts for 43.84\% of all errors, \texttt{TypeError} for 21.65\%, and \texttt{AssertionError} for 18.51\%. These top three error types collectively represent 84\% of all failures, indicating a concentrated failure signature. Chi-square goodness-of-fit testing against a uniform distribution confirmed highly non-uniform patterns ($\chi^2 = 37847.3$, $p < 10^{-300}$), with ``Other'' errors contributing 10.42\%. Notably, \texttt{SyntaxError} accounted for 0\% of failures, demonstrating that modern LLMs generate syntactically valid Python code consistently across all experimental conditions.

\begin{table}[t]
\centering
\caption{Overall error type distribution across all conditions.}
\label{tab:rq4_overall}
\begin{tabular}{l|c|c}
\toprule
\textbf{Error Type} & \textbf{Count} & \textbf{Percentage (\%)} \\
\midrule
AttributeError & 11,270 & 43.84 \\
TypeError & 5,566 & 21.65 \\
AssertionError & 4,759 & 18.51 \\
Other & 2,679 & 10.42 \\
ValueError & 702 & 2.73 \\
NameError & 339 & 1.32 \\
KeyError & 156 & 0.61 \\
FileNotFoundError & 89 & 0.35 \\
RuntimeError & 46 & 0.18 \\
ZeroDivisionError & 39 & 0.15 \\
IndexError & 38 & 0.15 \\
ImportError & 22 & 0.09 \\
SyntaxError & 0 & 0.00 \\
\bottomrule
\end{tabular}
\end{table}

\subsubsection{Dataset-Wise Comparison: Synthetic vs.\ Real-World Error Patterns}

\begin{figure*}[t]
\centering
\includegraphics[width=\columnwidth]{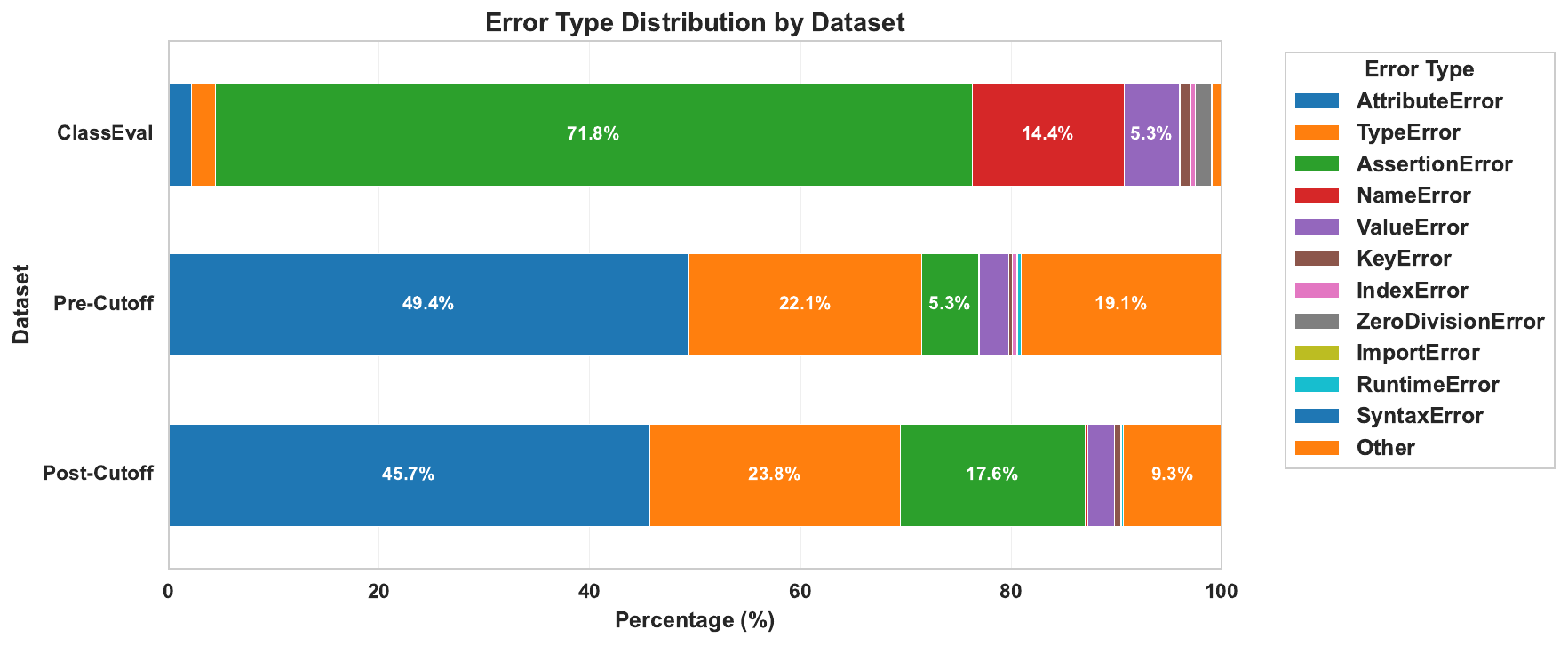}
\caption{Distribution of Error Types Across Datasets.}
\label{fig:error_distribution}
\end{figure*}

Error distributions differ fundamentally between synthetic benchmarks and real-world projects (Figure~\ref{fig:error_distribution}), with large effect sizes across all model-dataset comparisons (Table~\ref{tab:rq4_dataset}). All 21 pairwise comparisons showed statistically significant differences after FDR correction. ClassEval vs. pre-cutoff and ClassEval vs. post-cutoff comparisons yielded large effect sizes (Cramér's V: 0.59--0.85, all $p_{\text{FDR}} < 10^{-116}$), while pre-cutoff vs. post-cutoff comparisons showed smaller but still significant differences (Cramér's V: 0.22--0.28, all $p_{\text{FDR}} < 10^{-18}$). Statistical power exceeded 0.99 for all tests, ensuring reliable detection of effects.

\begin{table*}[t]
\centering
\caption{Dataset-wise error distribution comparison with global FDR correction (21 tests). All comparisons are significant after correction.}
\label{tab:rq4_dataset}
\begin{tabular}{l|l|c|c|c}
\toprule
\textbf{Model} & \textbf{Comparison} & \textbf{$p_{\text{raw}}$} & \textbf{$p_{\text{FDR}}$} & \textbf{Effect Size} \\
\midrule
\multirow{3}{*}{Codestral} & ClassEval vs Pre-Cutoff & 5.46e-144 & \textbf{1.04e-143} & Large \\
 & ClassEval vs Post-Cutoff & 3.10e-116 & \textbf{4.65e-116} & Large \\
 & Pre-Cutoff vs Post-Cutoff & 1.01e-28 & \textbf{1.32e-28} & Small \\
\hline
\multirow{3}{*}{Deepseek-V3} & ClassEval vs Pre-Cutoff & 5.91e-164 & \textbf{2.07e-163} & Large \\
 & ClassEval vs Post-Cutoff & 1.75e-142 & \textbf{3.07e-142} & Large \\
 & Pre-Cutoff vs Post-Cutoff & 3.23e-30 & \textbf{4.52e-30} & Small \\
\hline
\multirow{3}{*}{GPT-4.1} & ClassEval vs Pre-Cutoff & 1.01e-164 & \textbf{4.23e-164} & Large \\
 & ClassEval vs Post-Cutoff & 1.43e-136 & \textbf{2.32e-136} & Large \\
 & Pre-Cutoff vs Post-Cutoff & 3.89e-24 & \textbf{4.80e-24} & Small \\
\hline
\multirow{3}{*}{GPT-5} & ClassEval vs Pre-Cutoff & 2.17e-176 & \textbf{2.72e-175} & Large \\
 & ClassEval vs Post-Cutoff & 2.52e-157 & \textbf{7.56e-157} & Large \\
 & Pre-Cutoff vs Post-Cutoff & 5.45e-24 & \textbf{6.35e-24} & Small \\
\hline
\multirow{3}{*}{GPT-OSS} & ClassEval vs Pre-Cutoff & 1.76e-168 & \textbf{9.25e-168} & Large \\
 & ClassEval vs Post-Cutoff & 4.94e-145 & \textbf{1.04e-144} & Large \\
 & Pre-Cutoff vs Post-Cutoff & 3.77e-18 & \textbf{3.77e-18} & Small \\
\hline
\multirow{3}{*}{Llama 4 Maverick} & ClassEval vs Pre-Cutoff & 2.59e-176 & \textbf{2.72e-175} & Large \\
 & ClassEval vs Post-Cutoff & 4.71e-148 & \textbf{1.10e-147} & Large \\
 & Pre-Cutoff vs Post-Cutoff & 4.33e-21 & \textbf{4.79e-21} & Small \\
\hline
\multirow{3}{*}{Qwen Coder 2.5} & ClassEval vs Pre-Cutoff & 6.69e-176 & \textbf{4.68e-175} & Large \\
 & ClassEval vs Post-Cutoff & 3.04e-150 & \textbf{7.97e-150} & Large \\
 & Pre-Cutoff vs Post-Cutoff & 1.36e-19 & \textbf{1.42e-19} & Small \\
\bottomrule
\end{tabular}
\end{table*}

Detailed analysis of error type frequencies by dataset (Table~\ref{tab:rq4_dataset_breakdown}) reveals the source of these differences. ClassEval's error profile is dominated by \texttt{AssertionError} (71.80\%), reflecting the logic-focused nature of synthetic test suites, with \texttt{NameError} (14.44\%) as the second most common failure. On the other hand, real-world datasets show \texttt{AttributeError} dominance (49.41\% pre-cutoff, 45.67\% post-cutoff) and substantial \texttt{TypeError} frequencies (22.14\% pre-cutoff, 23.78\% post-cutoff), while \texttt{AssertionError} drops to 5.34\% (pre-cutoff) and 17.57\% (post-cutoff).

Quantifying these differences reveals striking patterns. ClassEval produces 66.47 percentage points (pp) more \texttt{AssertionError} than pre-cutoff and 54.24 pp more than post-cutoff, while showing 47.27 pp fewer \texttt{AttributeError} than pre-cutoff and 43.53 pp fewer than post-cutoff. \texttt{NameError} is 14.32 pp more common in ClassEval than pre-cutoff and 14.17 pp more than post-cutoff. Conversely, real-world code exhibits 19.80--21.45 pp more \texttt{TypeError} than ClassEval. These differences directly explain RQ1's performance gap: synthetic benchmarks test logical correctness through assertions, while real-world projects require correct object attribute access and type consistency---capabilities LLMs struggle to master.

\begin{table}[t]
\centering
\caption{Error type distributions by dataset (\%). ClassEval shows AssertionError dominance, while real-world datasets show AttributeError and TypeError prominence.}
\label{tab:rq4_dataset_breakdown}
\begin{tabular}{l|c|c|c}
\toprule
\textbf{Error Type} & \textbf{ClassEval} & \textbf{Pre-Cutoff} & \textbf{Post-Cutoff} \\
\midrule
AssertionError & 71.80 & 5.34 & 17.57 \\
AttributeError & 2.14 & 49.41 & 45.67 \\
TypeError & 2.33 & 22.14 & 23.78 \\
NameError & 14.44 & 0.12 & 0.27 \\
Other & 0.63 & 18.29 & 9.16 \\
ValueError & 5.30 & 2.81 & 2.53 \\
ZeroDivisionError & 1.51 & 0.14 & 0.00 \\
KeyError & 1.07 & 0.27 & 0.61 \\
IndexError & 0.44 & 0.42 & 0.01 \\
FileNotFoundError & 0.29 & 0.76 & 0.18 \\
RuntimeError & 0.05 & 0.30 & 0.20 \\
ImportError & 0.00 & 0.00 & 0.03 \\
SyntaxError & 0.00 & 0.00 & 0.00 \\
\bottomrule
\end{tabular}
\end{table}

The small but significant differences between pre-cutoff and post-cutoff (Cramér's V: 0.22--0.28) contrast with the negligible pass rate differences found in RQ1. While seen and unseen real-world code exhibit similar overall correctness, they show subtle error pattern variations. pre-cutoff code produces more ``Other'' errors (18.29\% vs.\ 9.16\%), suggesting idiosyncratic project-specific failures in older codebases, while post-cutoff shows slightly more \texttt{AssertionError} (17.57\% vs.\ 5.34\%), potentially reflecting evolution in testing practices toward more explicit assertions in recent projects.

\subsubsection{Docstring Impact on Error Patterns}

Similar to the docstrings' effect on overall pass rate (RQ2), they do not significantly affect error type distributions either. Of 28 pairwise comparisons (full vs.\ partial, full vs.\ no, across 7 models and 2 datasets), none achieved statistical significance after FDR correction (all $p_{\text{FDR}} > 0.05$). Effect sizes ranged from negligible to small (Cramér's V: 0.04--0.18), with most comparisons showing negligible effects. Power analysis revealed, on average, moderate detection capability (power: 0.19--0.92), suggesting that observed non-significance reflects a genuine absence of large effects.

This finding indicates that docstrings do not change error types even if, in isolated cases, they improve the pass rate. Models make the same categories of mistakes (primarily \texttt{AttributeError}, \texttt{TypeError}, and \texttt{AssertionError}) regardless of documentation completeness. Documentations affect all error types proportionally, rather than preventing or increasing specific failure modes selectively.

\subsubsection{RAG Error Substitution Mechanism}

\begin{figure*}[t]
\centering
\includegraphics[width=\columnwidth]{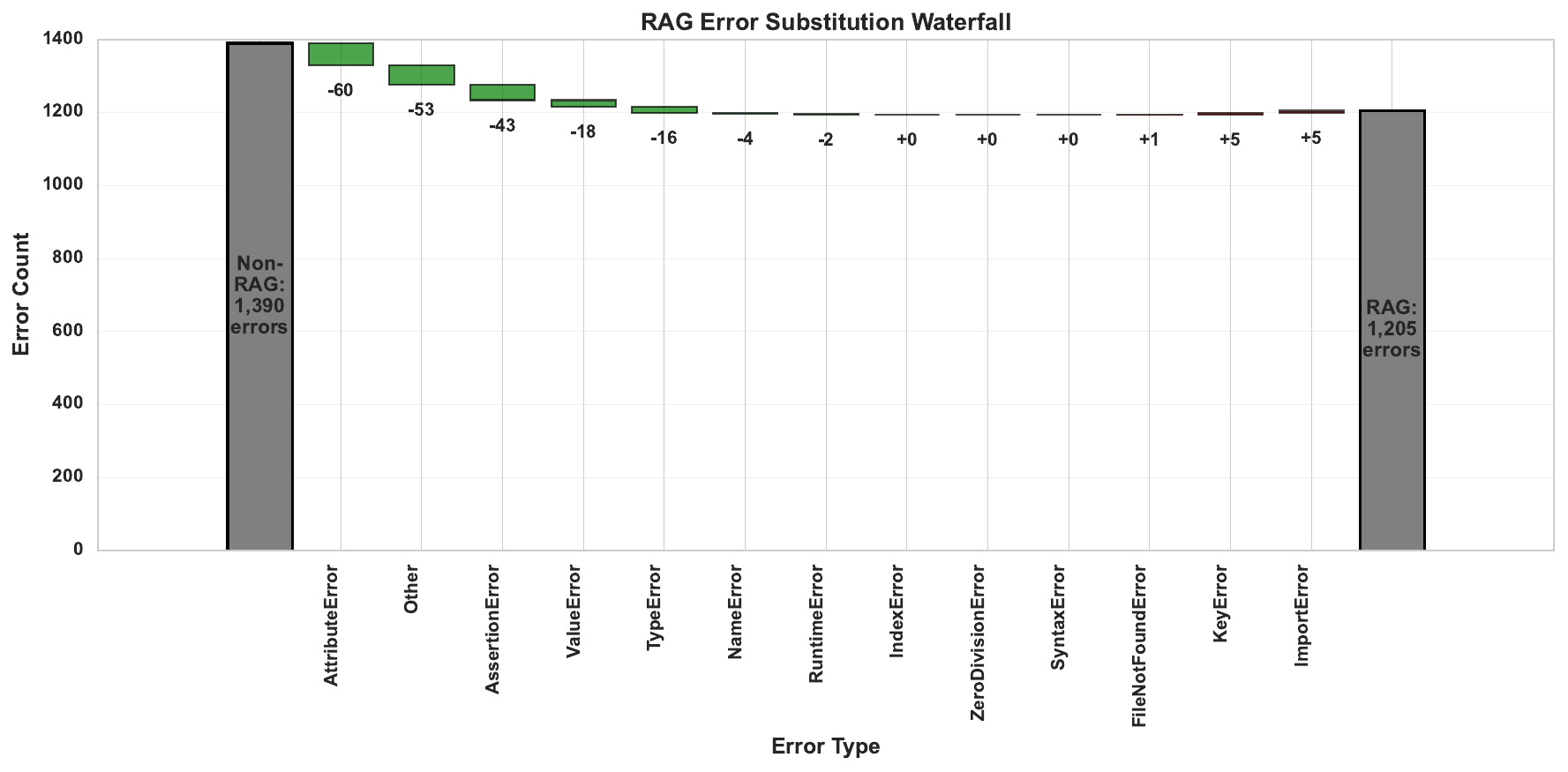}
\caption{Error Type Changes with RAG Implementation.}
\label{fig:rag_error_distribution}
\end{figure*}

An overall pattern of RAG's effect on error distributions is shown in Figure~\ref{fig:rag_error_distribution}. RAG not only improves pass rates with partial docstrings (RQ3) but also significantly alters error type distributions for three models: Deepseek-V3 ($\chi^2 = 20.68$, $p_{\text{FDR}} = 0.030$), GPT-5 ($\chi^2 = 16.62$, $p_{\text{FDR}} = 0.047$), and Llama 4 Maverick ($\chi^2 = 18.78$, $p_{\text{FDR}} = 0.047$), all with partial docstrings. Effect sizes are small (Cramér's V: 0.14--0.16) but statistically significant with adequate power (0.86--0.94). No significant differences emerged for full or no docstring conditions, consistent with RQ3's ``information gap hypothesis.''

Detailed analysis of error substitution patterns (Table~\ref{tab:rq4_rag_substitution}) reveals RAG's dual mechanism: substantial reductions in logic and object access errors, coupled with occasional increases in dependency-related failures. Aggregating across the three significant models, RAG with partial docstrings reduces 13.31\% of all errors, but the reduction is not uniform across error types.

\texttt{AttributeError} shows the largest absolute reduction (639 $\rightarrow$ 579, $-60$ instances), followed by ``Other'' errors (129 $\rightarrow$ 76, $-53$), \texttt{AssertionError} (233 $\rightarrow$ 190, $-43$), \texttt{ValueError} (46 $\rightarrow$ 28, $-18$), and \texttt{TypeError} (320 $\rightarrow$ 304, $-16$). These reductions suggest that retrieved examples help models understand correct object attribute access patterns, type specifications, and logical requirements, preventing the most common failure modes.

However, RAG introduces new failures in specific categories. \texttt{ImportError} increases from 2 to 7 instances (+5, +250\%), and \texttt{KeyError} from 11 to 16 (+5, +45\%). These increases reflect a substitution mechanism: retrieved examples contain dependencies or data structures absent in the target class, leading models to generate code that references unavailable imports or dictionary keys. The pattern is particularly pronounced in GPT-5, which shows a substantial increase in \texttt{ImportError} (0 $\rightarrow$ 7), and Deepseek-V3, which also exhibits a significant increase in \texttt{KeyError} (1 $\rightarrow$ 12).

Interestingly, while \texttt{AttributeError} and \texttt{TypeError} decrease in absolute count, their relative proportions increase (45.97\% $\rightarrow$ 48.05\% and 23.02\% $\rightarrow$ 25.23\%, respectively). This occurs because other error types decrease more substantially, making these harder-to-eliminate errors more prominent in the residual failure set. This suggests that \texttt{AttributeError} and \texttt{TypeError} represent fundamental challenges in class-level code generation that retrieval cannot fully address, as they require deep semantic understanding of object hierarchies and type systems that transcend pattern matching.

\begin{table*}[t]
\centering
\caption{RAG error substitution patterns for significant models with partial docstrings (aggregate across Deepseek-V3, GPT-5, Llama 4 Maverick). Negative differences indicate reductions, positive differences indicate increases.}
\label{tab:rq4_rag_substitution}
\resizebox{\columnwidth}{!}{
\begin{tabular}{l|c|c|c|c|c|c}
\toprule
\textbf{Error Type} & \textbf{Non-RAG} & \textbf{RAG} & \textbf{Absolute} & \textbf{Non-RAG} & \textbf{RAG} & \textbf{Percentage Point} \\
 & \textbf{Count} & \textbf{Count} & \textbf{Difference} & \textbf{\%} & \textbf{\%} & \textbf{Difference} \\
\midrule
\multicolumn{7}{l}{\textbf{\textit{Reductions}}} \\
AttributeError & 639 & 579 & $-60$ & 45.97 & 48.05 & +2.08 \\
Other & 129 & 76 & $-53$ & 9.28 & 6.31 & $-2.97$ \\
AssertionError & 233 & 190 & $-43$ & 16.76 & 15.77 & $-0.99$ \\
ValueError & 46 & 28 & $-18$ & 3.31 & 2.32 & $-0.99$ \\
TypeError & 320 & 304 & $-16$ & 23.02 & 25.23 & +2.21 \\
NameError & 6 & 2 & $-4$ & 0.43 & 0.17 & $-0.26$ \\
\midrule \midrule
\multicolumn{7}{l}{\textbf{\textit{Increases}}} \\
ImportError & 2 & 7 & \textbf{+5} & 0.14 & 0.58 & \textbf{+0.44} \\
KeyError & 11 & 16 & \textbf{+5} & 0.79 & 1.33 & \textbf{+0.54} \\
FileNotFoundError & 2 & 3 & +1 & 0.14 & 0.25 & +0.11 \\
\bottomrule
\end{tabular}}
\end{table*}

\subsection{Interpretation}
\label{subsec:rq4_interpretation}

The error analysis reveals three critical insights into LLM class-level code generation failures: a concentrated error signature dominated by object access and type consistency issues, fundamental differences between synthetic and real-world failure modes, and a RAG-induced error substitution mechanism. These findings provide actionable guidance for both practitioners and researchers seeking to improve code generation quality.

\subsubsection{The AttributeError-TypeError-AssertionError Triad}

The dominance of three error types, namely \texttt{AttributeError} (43.84\%), \texttt{TypeError} (21.65\%), and \texttt{AssertionError} (18.51\%), accounting for 84\% of all failures, reveals a consistent failure signature across models and conditions. This concentration indicates that LLM code generation struggles stem from specific, identifiable challenges rather than diffuse implementation errors.

\texttt{AttributeError} predominance directly links to RQ1's qualitative finding that real-world test suites involve complex type metadata checks and object hierarchies. LLMs generate plausible class structures but fail to correctly implement attribute access patterns, particularly in inheritance hierarchies and dynamic attribute resolution. This suggests that current training paradigms, which excel at surface-level syntax and common patterns, do not sufficiently capture deep object-oriented semantics.

\texttt{TypeError} frequency reflects challenges in maintaining type consistency across method calls, particularly in dynamically typed Python, where type annotations may be incomplete or absent. RQ2 showed that docstrings with explicit type information improve correctness by 1--3\%, but the persistence of \texttt{TypeError} even with full documentation suggests that models struggle to propagate type constraints through code generation, often mixing incompatible types or calling methods with incorrect argument types.

\texttt{AssertionError} prevalence indicates logical implementation failures where generated code produces incorrect values despite syntactic validity. The stark difference between ClassEval (71.80\%) and real-world datasets (5.34--17.57\%) demonstrates that synthetic benchmarks primarily test logical correctness through explicit assertions, while real-world failures manifest through object access and type issues before reaching assertion points.

The absence of \texttt{SyntaxError} (0.00\%) confirms that modern LLMs have mastered Python syntax completely. Even though this is opposite to the bug patterns in function-level LLM-generated code reported in a prior study~\citep{tambon2025bugs}, our finding validates LLMs' use for code generation tasks and shifts attention to semantic correctness---understanding object models, type systems, and logical specifications---as the remaining frontier.

\subsubsection{Synthetic vs.\ Real-World Error Type Divergence}

The large effect sizes (Cramér's V: 0.59--0.85) comparing ClassEval to real-world datasets definitively answer RQ1's open question about \textit{why} performance gaps exist. Synthetic benchmarks do not merely present easier versions of the same challenges; they represent fundamentally different problem spaces with distinct failure modes.

ClassEval's assertion-dominated error profile (71.80\% \texttt{AssertionError}) reflects its design philosophy: test logical correctness through input-output matching with simple equality checks. This aligns with the qualitative analysis in RQ1, which found ClassEval test suites use straightforward assertions like numerical comparisons and dictionary checks, requiring no external dependencies. LLMs perform well (83.67--88.51\% pass rate) because they can generate logically correct implementations for well-specified pure functions.

Real-world code's attribute-and-type-dominated profile (45--49\% \texttt{AttributeError}, 22--24\% \texttt{TypeError}) reflects genuine software engineering challenges: navigating complex object hierarchies, managing dependencies, and maintaining type consistency across diverse contexts. RQ1's qualitative analysis revealed that real-world test suites involve system dependencies (e.g., \texttt{numpy}, \texttt{MCPContext}) and metadata checks, leading to \texttt{AttributeError} when models misunderstand object structures. The 47 percentage point increase in \texttt{AttributeError} from ClassEval to pre-cutoff quantifies this fundamental shift.

The \texttt{NameError} pattern (14.44\% ClassEval vs.\ 0.12--0.27\% real-world) reveals another dimension of divergence. In ClassEval, models often fail to correctly reference variables or function names in self-contained implementations, while real-world code's lower \texttt{NameError} rate suggests that context from surrounding project code helps models resolve names correctly. This paradoxically indicates that isolated synthetic tasks may be harder for name resolution than contextualized real-world scenarios.

The small but significant pre-cutoff versus post-cutoff differences (Cramér's V: 0.22--0.28) warrant brief discussion despite their minor magnitude. While RQ1 found negligible pass rate differences between seen and unseen real-world code, error patterns show subtle temporal evolution. pre-cutoff code produces more ``Other'' errors (18.29\% vs.\ 9.16\%), suggesting older projects contain more idiosyncratic failure modes or deprecated patterns unfamiliar to models. post-cutoff's higher \texttt{AssertionError} (17.57\% vs.\ 5.34\%) may reflect modern testing practices favouring explicit assertions, though this requires further investigation.

\subsubsection{RAG Error Substitution: Trading Logic Errors for Dependency Failures}

RQ3 demonstrated that RAG improves pass rates by 4--7\% with partial docstrings, but the error distribution analysis reveals the \textit{mechanism}: RAG prevents logic and object access failures while introducing dependency-related errors. This substitution pattern is not merely error reduction but error transformation.

The substantial reductions in \texttt{AttributeError}, \texttt{AssertionError}, and \texttt{ValueError} indicate that retrieved examples provide concrete implementations that models can adapt, particularly for object attribute access patterns and logical requirements. When partial docstrings specify \textit{what} methods should do but not \textit{how}, retrieved similar implementations fill the gap, showing models correct attribute usage and logic structure.

However, the increases in \texttt{ImportError} and \texttt{KeyError} reveal RAG's limitation: retrieved examples may contain dependencies or data structures absent in the target context. Models blindly copy import statements or dictionary key accesses from examples without verifying availability in the target class. GPT-5's dramatic \texttt{ImportError} increase (0 $\rightarrow$ 7) suggests that this model particularly struggles to filter context-specific dependencies from retrieved examples, while Deepseek-V3's \texttt{KeyError} explosion (1 $\rightarrow$ 12) indicates difficulty distinguishing data structure patterns from concrete key names.

The relative proportion increases for \texttt{AttributeError} (+2.08 pp) and \texttt{TypeError} (+2.21 pp) despite absolute reductions, revealing these as ``hardcore'' errors resistant to retrieval-based mitigation. Even with concrete examples, models struggle to fully master object attribute navigation and type consistency, suggesting these require deeper semantic understanding beyond pattern matching.

It is worth mentioning that Llama 4 Maverick shows the most balanced substitution profile, reducing errors across the board with minimal dependency increases, suggesting architectural differences in how models integrate retrieved context. Future work should investigate these model-specific retrieval utilization strategies to optimize RAG effectiveness.

\subsubsection{Reflection on Previous RQs}
RQ4 cohesively integrates with prior research questions. RQ1 identified a performance gap between synthetic and real-world code (83.60--88.27\% vs.\ 24.59--33.60\% pass rates); RQ4 explains this through fundamentally different error types: ClassEval tests logical correctness through assertions, while real-world code requires object attribute mastery. RQ2 showed docstrings improve correctness by 1--3\% even though mostly statistically non-significant; RQ4 reveals they reduce all error types proportionally without changing failure modes, clarifying the documentation's mechanism. RQ3 demonstrated RAG's 4--7\% improvement with partial docs; RQ4 exposes the error substitution mechanism, trading logic failures for dependency errors, explaining both benefits and limitations.

\begin{tcolorbox}
    \textbf{RQ4 Findings:} LLM class-level code generation failures follow a consistent, concentrated signature dominated by object access and type errors. Synthetic benchmarks fundamentally mischaracterize these challenges by focusing on logical assertions. RAG offers promising but partial mitigation through error substitution.
\end{tcolorbox}

\section{Discussion}
\label{sec:discussion}

This study provides the first comprehensive evaluation of LLM performance on class-level code generation using real-world projects, revealing fundamental gaps between synthetic benchmark performance and practical capabilities. In this section, we provide a combined summary of all RQs and share actionable insights for practitioners and researchers.

\subsection{Key Findings}
\subsubsection{Performance Gap Between Synthetic and Real-World Code}

Our investigation of LLM performance across synthetic and real-world datasets revealed a sharp divide that questions current evaluation practices. All seven models achieved high pass rates on ClassEval, matching previous research findings. However, their performance dropped dramatically on real-world projects. The very small differences between pre-cutoff and post-cutoff show that models handle both seen and unseen real-world code equally well--but in both cases, the performance is poor for class-level code generation.

Our detailed analysis showed that this gap comes from different task types rather than harder problems. ClassEval test suites use simple equality checks with few external requirements, allowing models to achieve high pass rates through basic logic. Real-world test suites involve complex type checks, external system requirements, and integration between components--patterns consistent with genuine object-oriented implementation difficulty, though part of this gap may also reflect the absence of repository-level context.

\subsubsection{Docstrings as Incremental Improvement}

Our systematic test across three documentation levels (full, partial, no docstrings) revealed that complete docstrings may provide improvements in isolated cases. The pattern of benefits differed between time periods, with stronger effects on pre-cutoff data compared to post-cutoff. This suggests models rely more on documentation for familiar code patterns, where specifications help clarify how to implement things. However, comparing across time periods showed large differences between individual models. Some models performed better with post-cutoff data, while others performed better with pre-cutoff data. This indicates that docstring effectiveness depends on how models are built and what data they were trained on, rather than simply whether the code was seen during training.

\subsubsection{RAG's Information Gap Hypothesis}

Our retrieval pipeline showed a clear pattern with documentation completeness, supporting an ``information gap hypothesis" for this configuration: retrieval helps most when documentation is incomplete. This pattern shows how RAG works: when documentation explains what the code should do but doesn't show how to implement it (partial docstrings), retrieved examples fill this gap by providing actual code implementations.

Notably, GPT-5 performed worse with RAG, suggesting differences in how this model processes context. However, five other models with different designs (Codestral, Deepseek-V3, GPT-4.1, Llama 4 Maverick, Qwen Coder 2.5) all showed consistent positive results. This strengthens the applicability of the ``information gap" hypothesis across models, for the retrieval configuration studied here.

\subsubsection{Error Signatures and Substitution Mechanisms}

Error analysis revealed a focused pattern of failures: \texttt{AttributeError}, \texttt{TypeError}, and \texttt{AssertionError} make up 84\% of all failures. Notably, \texttt{SyntaxError} contributed 0\%, confirming that modern LLMs have fully learned Python syntax, making semantic correctness (understanding meaning) the main remaining challenge.

Dataset comparison revealed very different types of errors between synthetic and real-world code. ClassEval's error profile is mostly \texttt{AssertionError}, reflecting tests that check logic, while real-world datasets show mostly \texttt{AttributeError} and \texttt{TypeError}, reflecting real complexity in object-oriented programming.

RAG's impact on error distributions revealed a pattern of replacing one error with another rather than reducing all errors equally. Across the three models with strong RAG effects (Deepseek-V3, GPT-5, Llama 4 Maverick), RAG reduced \texttt{AttributeError}, \texttt{AssertionError}, and \texttt{ValueError}, showing that retrieved examples help prevent logic mistakes and problems accessing objects. However, RAG increased \texttt{ImportError} and \texttt{KeyError}, suggesting that retrieved examples contain dependencies or data structures not present in the target code, which models copy without checking if they fit.

\subsection{Class-Level Generation: The Critical Research Gap}
Class-level code generation occupies a crucial but understudied position in the granularity spectrum between well-researched function-level and rapidly advancing repository-level generation. Rather than representing an arbitrary choice of scope, class-level evaluation addresses a fundamental research gap that limits our understanding of LLM capabilities and constrains the effectiveness of repository-level approaches.

\subsubsection{The Granularity Progression Problem}
Current code generation research exhibits a striking discontinuity. Function-level generation has been extensively studied through benchmarks like HumanEval~\citep{chen2021evaluating}, MBPP~\citep{austin2021program}, and their variants, establishing a clear understanding of LLM capabilities for isolated, single-purpose code units. Repository-level generation has emerged as the frontier challenge, with benchmarks like SWE-bench~\citep{jimenez2023swe}, HumanEvo~\citep{zheng2024humanevo}, and CodeAgent~\citep{zhang2024codeagent} addressing complex multi-file scenarios.

However, this progression skips a critical intermediate step: \textit{Can LLMs reliably implement individual classes given clear specifications?} Repository-level tasks inherently require successful class implementation--SWE-bench issues often involve creating or modifying individual classes, CodeAgent tools ultimately generate class implementations, and environment-aware generation must ensure classes execute correctly across configurations. Yet existing evaluations cannot isolate whether failures stem from fundamental class-level implementation deficiencies or repository-level context management challenges.

\subsubsection{The Diagnostic Necessity}
Class-level evaluation serves a crucial diagnostic function that repository-level benchmarks cannot provide. When a repository-level approach fails, the failure could originate from multiple sources: inadequate context retrieval, poor dependency resolution, insufficient architectural scaffolding, or simply the inability to implement individual classes correctly. Without understanding baseline class-level capabilities, researchers cannot determine where to focus their architectural innovations.

For instance, if an LLM cannot correctly implement a class given a complete specification and relevant imports, then sophisticated repository-level context management becomes irrelevant--the fundamental implementation capability is missing. Conversely, if class-level implementation succeeds but repository-level tasks fail, this localizes the problem to context management and dependency resolution, guiding research toward more effective retrieval and architectural strategies.

\subsubsection{The Prerequisites for Repository-Level Success}
Repository-level generation approaches make implicit assumptions about class-level capabilities that have never been systematically validated. Consider the typical repository-level workflow:

\begin{enumerate} [leftmargin=*]
\item \textbf{Context Retrieval}: Gathering relevant code snippets, documentation, and dependency information
\item \textbf{Dependency Resolution}: Understanding cross-file relationships and import requirements  
\item \textbf{Code Implementation}: Actually generating the required class or function code
\item \textbf{Integration}: Ensuring generated code integrates correctly with the existing repository structure
\end{enumerate}

Current repository-level work focuses primarily on steps 1, 2, and 4--sophisticated retrieval mechanisms, dependency analysis, and integration strategies. However, step 3 receives minimal attention, with the implicit assumption that LLMs can successfully implement individual classes once provided with appropriate context.

Our findings challenge this assumption. If LLMs achieve only 25--34\% correctness on isolated class implementation (as our results demonstrate), then even perfect context management cannot overcome this fundamental limitation. Repository-level approaches may be optimizing the wrong problem--developing increasingly sophisticated context management for models that cannot reliably implement the basic building blocks.

\subsubsection{Isolating Core Capabilities from Architectural Scaffolding}
Class-level evaluation enables precise measurement of core LLM implementation capabilities independent of tool augmentation, architectural scaffolding, or sophisticated prompt engineering. This isolation serves several critical research purposes:

\textbf{Capability Assessment}: Determining the baseline implementation abilities that repository-level approaches can build upon, establishing realistic expectations for what context management and retrieval can achieve.

\textbf{Error Localization}: Identifying whether failures originate from fundamental implementation deficiencies (attribute resolution, type consistency, object-oriented reasoning) or context management issues, enabling targeted research efforts.

\textbf{Progress Measurement}: Providing a stable evaluation target that tracks improvements in core generation capabilities rather than confounding implementation ability with architectural sophistication.

\textbf{Transfer Understanding}: Establishing whether implementation patterns learned from repository context transfer to novel classes, or whether each class requires extensive contextual scaffolding.

\subsubsection{The Object-Oriented Foundation}
Classes represent the fundamental abstraction in object-oriented programming, encapsulating state, behaviour, and relationships in ways that individual functions cannot capture. Class-level generation requires understanding:

\begin{itemize}[leftmargin=*]
\item \textbf{State Management}: Maintaining object state across method calls and lifecycle events
\item \textbf{Method Interaction}: Ensuring methods work cohesively to implement class contracts
\item \textbf{Inheritance Hierarchies}: Correctly implementing parent-child relationships and method overriding  
\item \textbf{Interface Compliance}: Satisfying abstract method requirements and protocol adherence
\item \textbf{Encapsulation Principles}: Proper access control and information hiding
\end{itemize}

These capabilities form the foundation for larger software systems. A repository consists of interacting classes; understanding how well LLMs can implement individual classes is a prerequisite to understanding how they can compose multiple classes into coherent systems.

In summary, class-level code generation represents neither an arbitrary granularity choice nor competition with repository-level work. Instead, it addresses a critical research gap that constrains understanding of LLM capabilities and limits the effectiveness of more sophisticated approaches. By systematically evaluating class-level implementation independent of repository context, we provide the diagnostic foundation necessary for advancing the entire spectrum of code generation research.

\subsection{Broader Implications}

This study provides the first comprehensive evaluation of LLM performance on class-level code generation using real-world projects, revealing fundamental gaps between synthetic benchmark performance and practical capabilities. In this section, we discuss actionable insights for practitioners, researchers, and the broader software engineering community.

\subsubsection{The Simplicity-Complexity Paradox in Class Generation}

Our convergence with RepoClassBench~\citep{deshpande2024class} reveals a striking paradox that fundamentally challenges current assumptions about class-level code generation approaches. Despite vastly different methodological complexity (\ie our class skeletons versus their natural language descriptions with repository context), both approaches achieve nearly identical baseline performance (25--34\% vs. 24--34\%). This convergence exposes a critical insight: \textbf{specification richness and contextual sophistication do not improve fundamental class implementation capability}. Our findings demonstrate that simple structural specifications (class signatures and method skeletons) achieve the same baseline effectiveness as complex natural language descriptions combined with repository context. This convergence is suggestive rather than conclusive, given the two studies' non-comparable evaluation instruments and task setups. With that caveat, implementation-level factors such as attribute handling, type consistency, and object-oriented logic may represent a meaningful bottleneck alongside specification richness or repository context, rather than being fully resolved by either. LLMs understand \textit{what} to generate regardless of specification complexity; the challenge lies in generating \textit{correct implementations}. Furthermore, RepoClassBench's 40--54 percentage point improvements through RRR's extensive tool scaffolding demonstrate what becomes possible when core limitations are circumvented through architectural solutions. However, their need for iterative tool-based refinement, static analysis integration, and agent-based reasoning frameworks underscores that these improvements come through \textit{working around} rather than \textit{solving} the fundamental implementation bottleneck we identify. While their approach successfully enables repository-level class generation, it requires substantial computational overhead and system complexity.

For practitioners, this convergence suggests that effective class generation tools can be built using simpler approaches that focus on the core implementation problem rather than elaborate specification mechanisms. The path to improved performance lies in addressing implementation-specific weaknesses, such as attribute handling, type consistency, and object-oriented semantics, rather than enhancing specification richness or contextual sophistication.

\subsubsection{Rethinking Benchmark Design and Evaluation}

The large performance gap between synthetic and real-world code (57--62 percentage points) and fundamentally different error patterns reveal critical limitations in current evaluation practices. If synthetic benchmarks primarily test logical correctness through simple assertions while real-world code requires an understanding of object-oriented concepts, then benchmark leaderboards provide misleading assessments of practical capabilities.

Future benchmarks must preserve realistic object structures, dependencies, and testing practices rather than simplifying tasks to basic function-level problems. The complete absence of SyntaxError across all conditions demonstrates that correct syntax generation is solved, and research focus should shift entirely to semantic understanding and correctness. This necessitates evaluation metrics that measure understanding of meaning: correct object attribute usage, type consistency, and dependency management.

\subsubsection{Practical Deployment Strategies}

For practitioners deploying LLM-based code generation tools, our findings provide concrete guidance:

\textbf{Set Realistic Expectations}: Organizations should expect 25--34\% pass rates rather than synthetic benchmark success rates (84--89\%) for real-world class-level generation. This means LLM-generated code requires mandatory human review and comprehensive testing rather than being production-ready.

\textbf{Implement Adaptive RAG}: The retrieval pipeline evaluated here (GraphCodeBERT embeddings, top-k=2 full implementations) provides clear benefits (4–7\% improvement) specifically with partial documentation. Organizations using a similar retrieval-augmented setup should consider enabling it when documentation is incomplete, disabling it when documentation is comprehensive, and validate that RAG-suggested imports and dependencies exist in the target context.

\textbf{Focus Review on Dominant Error Types}: Since AttributeError and TypeError account for two-thirds of failures, code review processes should prioritize checking object attribute access patterns and type consistency. Static analysis tools targeting these specific error types could efficiently catch the majority of LLM-generated mistakes.

\subsubsection{Research Directions for Enhanced Code Generation}

Our concentrated error patterns reveal specific research directions that could substantially improve generation capabilities:

\textbf{Object-Oriented Semantic Learning}: The dominance of AttributeError (43.84\%) indicates that current training inadequately captures object-oriented concepts. Future research should investigate training objectives specifically targeting attribute resolution, inheritance relationships, and dynamic attribute handling. Graph neural network architectures that explicitly model class hierarchies may better capture object relationships than transformer architectures alone.

\textbf{Type-Aware Generation Architectures}: The persistent TypeError challenge (21.65\%) motivates research into generation architectures that maintain explicit type information during generation, enforcing consistency requirements. Integration with gradual typing systems could provide training signals for type-correct generation.

\textbf{Error-Specific Training Curricula}: The concentrated three-error signature (AttributeError, TypeError, AssertionError accounting for 84\% of failures) suggests targeted training curricula could efficiently improve generation quality. Curriculum learning approaches that progressively increase object-oriented complexity may better develop semantic understanding than uniform dataset training.

\subsubsection{Methodological Contributions and Future Evaluation}

This work establishes methodological advances for code generation research:

\textbf{Real-World Evaluation Paradigm}: \textit{RealClassEval} demonstrates the feasibility and necessity of real-world code generation evaluation. The field requires systematic development of large-scale, diverse real-world benchmarks that preserve the complexity and authenticity missing from synthetic evaluations.

\textbf{Error-Focused Analysis Framework}: Moving beyond aggregate pass rates to detailed error type analysis reveals failure mechanisms rather than just failure rates. The RAG error substitution pattern—logic failures replaced by dependency errors—provides critical insights that would remain invisible in pass-rate-only evaluation.

\textbf{Temporal Evaluation Strategy}: The pre-cutoff/post-cutoff temporal split provides a rigorous approach to data contamination concerns while enabling systematic study of generalization patterns.

As LLM-based development tools become increasingly prevalent in software engineering workflows, rigorous evaluation on realistic tasks becomes essential. This work demonstrates that synthetic benchmarks provide false confidence about practical capabilities, while systematic real-world evaluation reveals both current limitations and specific paths forward. The class-level granularity investigated here provides the diagnostic foundation necessary for advancing both fundamental LLM capabilities and sophisticated repository-level approaches, ultimately contributing to more reliable and effective automated programming assistance.

\section{Limitations and Future Work}
\label{sec:limitations}

While this study provides valuable insights into LLM class-level code generation, several limitations warrant discussion and motivate future research.

\subsection{Limitations}

\subsubsection{Sample Size}

The most significant limitation concerns dataset size and the resulting small to negligible effect sizes in the majority of cases. Our study evaluated 200 classes per \textit{RealClassEval} version, totalling 400 real-world classes compared to ClassEval's 100 synthetic tasks. While each split (pre-cutoff and post-cutoff) of \textit{RealClassEval} is 2 times larger than the established ClassEval benchmark and represents the first carefully collected real-world class-level dataset, it remains relatively small by machine learning evaluation standards. This size limitation appears in consistently small effect sizes across research questions despite statistical significance. Docstring effects and RAG effects showed negligible practical differences, and error distribution comparisons between pre-cutoff and post-cutoff showed small effects compared to the synthetic-real divide. These small effect sizes reflect three factors: genuine small practical effects, high variability between classes, and limited statistical power to detect small-to-medium effects with 200 samples.

The limited size stems from practical constraints. Generating and testing over 16,000 classes across seven LLMs, two dataset versions, three docstring conditions, and two retrieval settings required substantial computational resources. At current API rates of different LLMs' API service providers, the total generation cost, including trial and error, was approximately \$1000 for API calls alone, plus over 500 CPU hours for test execution. Each additional class requires LLM generation, test suite generation, and test execution, increasing costs directly with size. Expanding to 1,000+ classes per version would require proportional resource scaling, exceeding typical academic research budgets.

However, our sample size is sufficient for establishing fundamental patterns. The large performance gap between synthetic and real-world code and the error type differences are robust with clear practical significance. The small effect sizes for docstrings and RAG reflect genuine small benefits rather than merely insufficient power, as evidenced by consistency across multiple models, statistically significant findings in key comparisons, and coherent patterns. Critically, this study provides the first evidence from real-world data on class-level code generation, establishing baselines and patterns that synthetic benchmarks alone cannot reveal. Future work with larger-scale real-world datasets is essential for more conclusive findings, particularly for detecting smaller effects and enabling more detailed analyses.

\subsubsection{Model Selection and Recency}

Our evaluation included seven state-of-the-art models as of October 2025, representing diverse architectures and capabilities. However, the rapid pace of LLM development means newer models may show different patterns. The consistency of findings across all seven models--particularly the performance gap between synthetic and real-world code and error type dominance--suggests robust patterns likely to persist in future models. Additionally, all evaluated models are transformer-based architectures, and alternative architectures like graph neural networks explicitly modeling code structure may show different strengths, particularly for understanding object-oriented concepts.

\subsubsection{Task Scope: Class-Level vs. Other Granularities}

This study focuses exclusively on class-level code generation, where models must implement complete classes with multiple methods given only skeletons. Findings may not generalize to other granularities: single-function generation may show higher success rates due to a simpler context, while generating entire projects likely shows even larger challenges due to cross-file dependencies. The choice of class-level granularity reflects realistic software engineering tasks and enables controlled evaluation, but complete understanding requires evaluation across multiple granularities. The error pattern insights may not generalize to other object-oriented granularities~\citep{tambon2025bugs} and require validation.

\subsubsection{Test Suite Generation Limitations}

We relied on Pynguin for automatic test suite generation, using its DynaMOSA algorithm to achieve high branch coverage. While this approach enables objective correctness assessment without manual test writing, automatically generated test suites have limitations: they primarily assess that code executes without exceptions, but may not validate whether methods compute correct results; some classes may have low-coverage test suites missing errors in uncovered paths. This limitation is asymmetric across our two benchmark types: PYNGUIN achieves 93.2\% average branch coverage on ClassEval but only 58.7\% on \textit{RealClassEval}, meaning functional errors in uncovered branches are somewhat more likely to go undetected on \textit{RealClassEval} than on ClassEval.
Section~\ref{subsec:coverage_analysis} reports a manual test-quality assessment (n=78) confirming that generated test suites are overwhelmingly assertion-based (88.5\%) rather than relying solely on implicit exception checks. This addresses whether tests contain explicit checks at all, but it does not establish that those assertions accurately capture the intended semantics of the original classes, nor does it measure the fault-detection capability (mutation adequacy) of the generated suites. Semantic and fault-detection adequacy of the automatically generated oracle therefore remains an open limitation of this study, one that manual test authoring or mutation-testing-based validation could help address in future work.

\subsubsection{Prompt Engineering and Generation Settings}

We used a standardized prompt across all models with consistent generation settings to ensure fair comparison, but this does not reflect optimal prompting strategies for each model. Some models may benefit from model-specific prompt formats, few-shot examples~\citep{yuen2025prompting}, or interactive refinement strategies. However, the finding that all seven models show similar error patterns suggests that fundamental capabilities rather than prompt optimization drive performance. The goal was to assess base capabilities under standardized conditions rather than optimizing each model individually, reflecting realistic deployment scenarios.

\subsubsection{Seen versus Unseen Benchmark Shelf Life}
Our pre-cutoff (nominally seen) versus post-cutoff (nominally unseen) split is intended to approximate corpus-level exposure, but it rests on assumptions worth stating explicitly. CodeSearchNet, the source corpus underlying our pre-cutoff sampling, is a function-level dataset of (comment, code) pairs, not a corpus of complete classes; even where a model's training data overlapped with CodeSearchNet, it would have encountered the constituent functions individually, not the class-level artifacts reconstructed for this study. Our seen versus unseen labelling therefore reflects exposure to related code patterns at a coarser granularity than the artifacts we evaluate, and should not be read as confirmed exposure to the specific classes under test. In the other direction, we cannot definitively verify that post-cutoff data is unseen: models may have encountered mirrors, forks, or near-duplicate code, and for closed models, stated knowledge cutoffs do not fully rule out post-training exposure, distillation from other models, or benchmark contamination through indirect channels. The small, largely non-significant differences we observe between pre-cutoff and post-cutoff performance are consistent with corpus-level familiarity not meaningfully improving real-world class-level generation, but given the above, this should be treated as suggestive rather than a definitive test of memorization.

Separately, temporal-split benchmarks like \textit{RealClassEval} have a limited shelf life by construction: repositories that are unseen at the time of dataset creation will eventually be scraped into future models' training corpora as time passes. \textit{RealClassEval}'s post-cutoff half should therefore be understood as unseen relative to the models evaluated in this study, not as a permanently unseen reference set. Reusing this benchmark to evaluate future models will require periodically refreshing the post-cutoff sample with newer repositories to preserve its intended function.

\subsubsection{Generalization Beyond Python}

This is a Python-only study, as suggested by the title of the paper. Therefore, all evaluated code is Python-based, drawn from open-source GitHub repositories. Python's dynamic typing likely shapes our error-type findings; the dominance of \texttt{AttributeError} and \texttt{TypeError} among generation failures may reflect the absence of compile-time type checking rather than a language-agnostic property of class-level generation difficulty, and statically typed languages could exhibit a different error-type profile entirely. Closed-source or enterprise code may also follow conventions or domain-specific patterns underrepresented in open-source training data, and the synthetic-versus-real-world performance gap we report may not transfer directly to those settings. This makes a multi-language and closed-source evaluation an important direction for future work.

\subsubsection{Evaluation Without Repository Context}

\textit{RealClassEval} evaluates class implementation in isolation from the surrounding repository: dependencies, sibling classes, and project-level context available to a developer (or a repository-aware tool) are not provided to the model. This is a deliberate scope choice given that we aim to measure class-level implementation capability given a skeleton and docstrings, isolated from confounding effects of context availability. But this also means part of the synthetic-versus-real-world performance gap reported in this paper likely reflects the absence of contextual information rather than a pure deficit in object-oriented reasoning. Real-world deployment scenarios that supply repository context (e.g., via retrieval-augmented pipelines, as explored in RQ3, or repository-aware coding agents) may narrow this gap; our results should be interpreted as a lower bound on context-free class implementation capability rather than a ceiling on real-world task performance more broadly.

Despite these limitations, this study establishes fundamental patterns in LLM class-level code generation, provides the first real-world evaluation baseline, and illuminates specific research directions for advancing code synthesis capabilities. The limitations identified here directly motivate ongoing work developing large-scale, diverse, quality-controlled real-world benchmarks to enable more conclusive future studies.

\subsection{Future Work}

The field urgently requires large-scale, diverse, real-world code generation benchmarks to enable definitive evaluation and advance beyond synthetic task limitations. An ideal benchmark would include:

\begin{itemize}[leftmargin=*]
\item \textbf{Scale}: 1,000+ classes per temporal category (pre-cutoff, post-cutoff), enabling detection of small-to-medium effect sizes with high power ($>0.80$ for Cohen's $d = 0.2$) and supporting subgroup analyses by domain, complexity, or project characteristics.

\item \textbf{Diversity}: Classes spanning multiple domains (web development, data science, systems programming, scientific computing), programming paradigms (object-oriented, functional, procedural mixtures), and complexity levels (simple data structures to complex business logic), ensuring generalization beyond specific niches.

\item \textbf{Quality Control}: Systematic filtering for engineered projects with mature testing practices, automated test suite validation (ensuring tests actually execute and provide coverage), and manual spot-checking to verify task quality and remove problematic edge cases.

\item \textbf{Metadata Richness}: Annotations for class complexity (cyclomatic complexity, coupling metrics), domain categories, required dependencies, and testing characteristics (unit vs.\ integration, coverage levels), enabling fine-grained analysis of where models succeed or fail.

\item \textbf{Multi-Language Support}: Extension beyond Python to include Java, JavaScript, C++, and other languages, assessing whether findings generalize across language ecosystems or reflect Python-specific patterns.
\end{itemize}

Developing such a benchmark represents a substantial research contribution in itself, requiring substantial curation effort and computational costs. The insights from this study--particularly the importance of preserving realistic object hierarchies, dependencies, and testing practices--provide design principles for next-generation benchmarks.

\section{Threats to Validity}
\label{sec:threats}

We discuss threats to validity following established guidelines for empirical software engineering research, addressing internal validity (confounding factors), external validity (generalizability), and construct validity (measurement accuracy).

\subsection{Internal Validity}

Internal validity concerns whether observed effects genuinely result from our experimental design rather than confounding variables. We employed several controls to ensure valid findings. We used identical prompts across all models and consistent generation settings to control for prompt-induced variance. All test suites were generated using Pynguin with an identical configuration, ensuring consistent evaluation across classes. We used random sampling for class selection, reducing selection bias toward specific project types or domains. For RQ2 and RQ3, we employed within-subjects designs where each class was evaluated under all docstring conditions and both retrieval settings, controlling for snippet-specific difficulty.

\subsection{External Validity}

External validity concerns generalizability beyond our specific experimental setting. Several factors limit generalizability. All evaluated code is Python-based, and findings may not generalize to statically typed languages like Java or C++, which would likely show fewer type errors due to compile-time checking but potentially more complex type hierarchy challenges. All data derives from open-source GitHub repositories, potentially limiting generalizability to proprietary enterprise codebases, which may follow different conventions, have more comprehensive documentation, or employ domain-specific patterns unfamiliar to models trained predominantly on open-source data.

\subsection{Construct Validity}

Construct validity concerns whether measurements accurately capture intended constructs. We use pass rate as the primary correctness metric, which has limitations: it depends entirely on test suite quality, measures behavioural rather than structural correctness, and doesn't distinguish between semantically equivalent implementations. However, pass rate is standard in code generation research~\citep{yeo2024framework,hendrycks2021measuring,yuen2025prompting}, provides an objective functional correctness measurement through execution, and is validated across multiple test cases per class. Our qualitative analysis examined test suites and failure modes, confirming that pass rates meaningfully distinguish correct implementations from failures.

A related threat concerns the reliability of our reported coverage figures (93.2\% ClassEval, 58.7\% \textit{RealClassEval}) as absolute measurements. A reproducibility check traced our ClassEval coverage shortfall relative to the 98.2\% figure ClassEval's own authors report~\citep{du2023classeval} to unresolved third-party dependencies in our measurement environment, which is a common artifact when re-executing real-world code at scale, and one likely present in the \textit{RealClassEval} figure as well. We retain the originally reported values for consistency with our experimental logs and treat both as conservative estimates.

\section{Conclusion}
\label{sec:conclusion}

This work provides the first comprehensive evaluation of large language model performance on class-level code generation using real-world projects, revealing fundamental limitations masked by synthetic benchmark evaluation. Through systematic analysis classes generated by seven state-of-the-art models, we demonstrate that current LLMs achieve 84--89\% correctness on synthetic benchmarks but only 25--34\% on real-world code--a 53--62 percentage point gap. This performance drop stems from fundamentally different error patterns rather than just harder problems: synthetic benchmarks test logical correctness through simple checks, while real-world code exercises more complex object-oriented patterns, a difficulty not fully separable from the absence of repository-level context in our evaluation design Real-world failures are dominated by \texttt{AttributeError}s and \texttt{TypeError}s, accounting for two-thirds of all failures, while synthetic benchmarks show mostly logic-induced errors. Modern LLMs have completely mastered Python syntax but struggle with object-oriented semantics. Models achieve nearly identical pass rates on seen and unseen real-world code, which we interpret as evidence against corpus-level familiarity being a major driver of real-world performance. As discussed in Section~\ref{sec:limitations}, however, this should not be read as a definitive test of memorization, given the function-level granularity of our seen-data corpus. This challenges current evaluation practices: synthetic benchmark performance does not accurately represent real-world capabilities, and organizations should expect much lower success rates when deploying LLM-based code generation tools.

Our studies reveal that documentation and retrieval provide complementary but modest improvements, with deployment of a similar retrieval setup likely requiring adaptive strategies based on documentation completeness. Beyond these immediate findings, this work addresses a critical gap in the code generation research landscape: understanding class-level capabilities as the essential foundation for repository-level approaches. The concentrated error patterns point to specific research directions: developing training objectives targeting attribute resolution and type consistency, creating context-aware retrieval systems that filter incompatible dependencies, and building large-scale real-world benchmarks spanning multiple languages and domains. 

As LLM-based development tools become increasingly common in software engineering and repository-level approaches advance rapidly, a systematic understanding of class-level capabilities becomes essential. This work establishes that synthetic benchmarks provide false confidence about fundamental implementation abilities, while class-level evaluation on real-world code reveals both current limitations and specific paths forward. The diagnostic foundation provided here enables more effective development of repository-level approaches and more reliable deployment of automated programming assistance in production environments. We make our data and evaluation framework publicly available to facilitate continued research in this critical area.

\section*{Acknowledgments}
We used AI-enabled tools (Claude and Grammarly) to assist with manuscript revisions. All AI-assisted content was reviewed and verified for accuracy by the first author.

\section*{Declarations}

\subsection*{Funding}
This research was supported by the \textbf{Natural Sciences and Engineering Research Council of Canada (NSERC)} under Grant Number: \textit{CREATE 555406-2021}.

\subsection*{Conflict of Interest}
\textbf{Emad Shihab} has ongoing collaborations with EMSE Editorial Board members: \emph{Foutse Khomh}, \emph{Sarah Nadi}, \emph{Mei Nagappan}, and \emph{Alexander Serebrenik}.

\subsection*{Ethical Approval}
This article does not contain any studies with human participants or animals performed by any of the authors.

\subsection*{Informed Consent}
Not applicable.

\subsection*{Author Contributions}
\textbf{Musfiqur Rahman:} Conceptualization, Methodology, Implementation, Formal analysis, Investigation, Data curation, Writing - original draft, Writing - review \& editing, Visualization. \textbf{SayedHassan Khatoonabadi:} Conceptualization, Methodology. \textbf{Emad Shihab:} Conceptualization, Writing - review, Supervision, Project administration, Funding acquisition.

\phantomsection
\subsection*{Data Availability Statement} \label{sec:replication-package}
The datasets generated and analyzed during the current study, along with all code and analysis scripts, are publicly available in our replication package at: \url{https://github.com/mrsumitbd/RealClassEval-Replication}.

\subsection*{Clinical Trial Number}
Not applicable.

\bibliographystyle{spbasic}      
\bibliography{references}   

\appendix

\section{Appendix}
\subsection{Repository filtering funnel} \label{app:filtering_funnel}

\begin{table}[h]
\centering
\caption{Repository filtering funnel for Pre-Cutoff and Post-Cutoff datasets.}
\label{tab:filtering_funnel}
\begin{tabular}{l|c|c}
\toprule
\textbf{Filtering Criterion} & \textbf{Pre-Cutoff} & \textbf{Post-Cutoff} \\
\midrule
Initial Pool & $13,590$ & $91,048$ \\
Filtered on Project Size and Activity & $5,348$ & $13,197$ \\
Filtered on License Declaration & $4,021$ & $4,292$ \\
Filtered on Code Quality Indicators & $2,970$ & $1,735$ \\
\bottomrule
\end{tabular}
\end{table}

\subsection{Full Per-Model Statistical Tables}
\label{app:full_tables}

The following table provides the full per-model breakdown underlying the condensed summary Tables~\ref{tab:rq1_perclass} and \ref{tab:rq1_pooled} in Section~\ref{sec:rq1}.

\begin{table}[h]
\centering
\caption{Per-Class Pass Rate Analysis (full per-model breakdown). \textbf{Bold} entries represent statistical significance.}
\label{tab:rq1_perclass_full}
\resizebox{\columnwidth}{!}{
\begin{tabular}{l|l|l|c|l}
\toprule
\textbf{Model} & \textbf{Comparison} & \textbf{Mean Diff (95\% CI)} & \textbf{\makecell{FDR-Adjusted $p$ for\\Mann-Whitney U Tests}} & \textbf{Effect Size}\\
\midrule
Codestral & ClassEval vs. Pre-Cutoff & 46.93 [39.46, 54.93] & \textbf{1.0233e-10} & Medium \\
& ClassEval vs. Post-Cutoff & 51.50 [44.36, 58.63] & \textbf{1.4623e-16} & Large \\
& Pre-Cutoff vs. Post-Cutoff & 4.57 [-4.32, 13.74] & 0.4623 & Negligible \\
\hline
Deepseek-V3 & ClassEval vs. Pre-Cutoff & 50.50 [42.82, 57.75] & \textbf{1.9786e-11} & Medium \\
& ClassEval vs. Post-Cutoff & 49.38 [42.82, 56.03] & \textbf{3.5676e-14} & Large \\
& Pre-Cutoff vs. Post-Cutoff & -1.12 [-9.49, 7.59] & 0.6604 & Negligible \\
\hline
GPT-4.1 & ClassEval vs. Pre-Cutoff & 59.45 [52.51, 65.89] & \textbf{6.5615e-19} & Large \\
& ClassEval vs. Post-Cutoff & 53.17 [47.17, 59.09] & \textbf{1.7335e-19} & Large \\
& Pre-Cutoff vs. Post-Cutoff & -6.28 [-13.98, 2.07] & 0.0613 & Negligible \\
\hline
GPT-5 & ClassEval vs. Pre-Cutoff & 61.24 [54.63, 68.16] & \textbf{1.4079e-20} & Large \\
& ClassEval vs. Post-Cutoff & 57.05 [49.81, 63.58] & \textbf{4.7559e-18} & Large \\
& Pre-Cutoff vs. Post-Cutoff & -4.18 [-12.33, 3.59] & 0.3013 & Negligible \\
\hline
GPT-OSS & ClassEval vs. Pre-Cutoff & 58.71 [51.48, 65.82] & \textbf{4.7559e-18} & Large \\
& ClassEval vs. Post-Cutoff & 49.66 [42.19, 56.93] & \textbf{7.1164e-16} & Large \\
& Pre-Cutoff vs. Post-Cutoff & -9.05 [-17.54, -0.46] & \textbf{0.0188} & Negligible \\
\hline
Llama 4 Maverick & ClassEval vs. Pre-Cutoff & 54.25 [46.70, 61.57] & \textbf{4.4041e-16} & Large \\
& ClassEval vs. Post-Cutoff & 49.68 [42.60, 56.61] & \textbf{7.1164e-16} & Large \\
& Pre-Cutoff vs. Post-Cutoff & -4.57 [-12.92, 3.78] & 0.1355 & Negligible \\
\hline
Qwen Coder 2.5 & ClassEval vs. Pre-Cutoff & 50.02 [42.22, 57.07] & \textbf{1.9965e-13} & Large \\
& ClassEval vs. Post-Cutoff & 43.96 [36.70, 51.25] & \textbf{7.5609e-12} & Medium \\
& Pre-Cutoff vs. Post-Cutoff & -6.07 [-14.40, 2.82] & 0.1200 & Negligible \\
\bottomrule
\end{tabular}
}
\end{table}

\begin{table}[h]
\centering
\caption{Pooled Chi-Square Analysis (full per-model breakdown).}
\label{tab:rq1_pooled_full}
\resizebox{\columnwidth}{!}{
\begin{tabular}{l|l|c|c|c}
\toprule
\textbf{Model} & \textbf{Comparison} & \textbf{Mean Diff. (95\% CI)} & \textbf{FDR-Adjusted $p$} & \textbf{Effect Size} \\
\midrule
\multirow{3}{*}{Codestral} & ClassEval vs. Pre-Cutoff & 51.75 [47.28, 56.83] & \textbf{1.9032e-106} & Medium \\
& ClassEval vs. Post-Cutoff & 57.37 [53.22, 61.24] & \textbf{1.7756e-168} & Large \\
& Pre-Cutoff vs. Post-Cutoff & 5.62 [-0.32, 11.33] & 0.1034 & Negligible \\
\hline
\multirow{3}{*}{Deepseek-V3} & ClassEval vs. Pre-Cutoff & 54.44 [49.81, 59.06] & \textbf{4.7305e-126} & Medium \\
& ClassEval vs. Post-Cutoff & 55.99 [52.14, 59.88] & \textbf{7.6134e-174} & Large \\
& Pre-Cutoff vs. Post-Cutoff & 1.55 [-3.73, 7.40] & 0.7654 & Negligible \\
\hline
\multirow{3}{*}{GPT-4.1} & ClassEval vs. Pre-Cutoff & 58.14 [52.86, 62.83] & \textbf{2.3701e-152} & Large \\
& ClassEval vs. Post-Cutoff & 57.43 [53.83, 61.24] & \textbf{1.8167e-193} & Large \\
& Pre-Cutoff vs. Post-Cutoff & -0.71 [-6.43, 5.09] & 0.8587 & Negligible \\
\hline
\multirow{3}{*}{GPT-5} & ClassEval vs. Pre-Cutoff & 60.36 [55.90, 65.02] & \textbf{7.4220e-155} & Large \\
& ClassEval vs. Post-Cutoff & 58.06 [53.93, 61.48] & \textbf{1.0034e-191} & Large \\
& Pre-Cutoff vs. Post-Cutoff & -2.29 [-7.94, 3.10] & 0.5707 & Negligible \\
\hline
\multirow{3}{*}{GPT-OSS} & ClassEval vs. Pre-Cutoff & 62.17 [57.34, 66.66] & \textbf{2.2409e-149} & Large \\
& ClassEval vs. Post-Cutoff & 58.28 [54.47, 62.06] & \textbf{7.6322e-179} & Large \\
& Pre-Cutoff vs. Post-Cutoff & -3.89 [-9.75, 1.57] & 0.2758 & Negligible \\
\hline
\multirow{3}{*}{Llama 4 Maverick} & ClassEval vs. Pre-Cutoff & 56.38 [51.37, 61.21] & \textbf{3.4048e-123} & Medium \\
& ClassEval vs. Post-Cutoff & 57.25 [53.33, 60.92] & \textbf{2.3489e-174} & Large \\
& Pre-Cutoff vs. Post-Cutoff & 0.87 [-4.96, 6.84] & 0.8587 & Negligible \\
\hline
\multirow{3}{*}{Qwen Coder 2.5} & ClassEval vs. Pre-Cutoff & 52.85 [47.58, 57.67] & \textbf{7.2034e-106} & Medium \\
& ClassEval vs. Post-Cutoff & 54.20 [50.31, 58.12] & \textbf{1.2561e-152} & Large \\
& Pre-Cutoff vs. Post-Cutoff & 1.35 [-4.57, 7.58] & 0.7779 & Negligible \\
\bottomrule
\end{tabular}
}
\end{table}

\end{document}